\documentclass[11pt, a4paper, oneside]{article} 
\usepackage[a4paper, left=19.1mm, right=19.1mm, top=25.4mm, bottom=25.4mm]{geometry}   
 
\usepackage{chemformula} % Formula subscripts using \ch{}
\usepackage[T1]{fontenc} % Use modern font encodings
\usepackage{subcaption}
\usepackage{siunitx}
\usepackage{multirow}  
\usepackage{array}
\usepackage{booktabs}
\usepackage{comment}
\usepackage[super, round, comma]{natbib}
\usepackage{hyperref}   
\usepackage{setspace}
\usepackage{upgreek}
\usepackage{enumitem}   
\hypersetup{colorlinks=true,linkcolor=blue,filecolor=blue,citecolor=blue,urlcolor=blue}

\makeatletter
\providecommand\add@text{}
\newcommand\tagaddtext[1]{%
	\gdef\add@text{#1\gdef\add@text{}}}% 
\renewcommand\tagform@[1]{%
	\maketag@@@{\llap{\add@text\quad}(\ignorespaces#1\unskip\@@italiccorr)}%
}
\makeatother

\makeatletter
\newcommand{\mathleft}{\@fleqntrue\@mathmargin0pt}
\newcommand{\mathcenter}{\@fleqnfalse}
\makeatother

\begin{document}
\doublespacing
	\begin{titlepage}
	\begin{center}
		\vspace*{0.5cm}
		\vspace*{0.5cm}
		{\huge \textbf{The influence of silicon on the formation and transformation of corrosion products}}\\
		\vspace*{0.5cm}
		{\large Fabio E. Furcas\textsuperscript{$\dagger$}, Shishir Mundra\textsuperscript{$\dagger$}, Barbara Lothenbach\textsuperscript{$\mathsection$}, Camelia N. Borca\textsuperscript{$\ddagger$}, Thomas Huthwelker\textsuperscript{$\ddagger$}, Ueli M. Angst\textsuperscript{$\dagger *$}}\\
		\vspace*{0.5cm}
		{\large \textsuperscript{$\dagger$}Institute for Building Materials, ETH Z\"{u}rich, 8093, Z\"{u}rich, Switzerland}\\
		{\large \textsuperscript{$\mathsection$}Empa Concrete \& Asphalt Labortory, 8600 D\"{u}bendorf, Switzerland}\\
		{\large \textsuperscript{$\ddagger$}Swiss Light Source, Paul Scherrer Institut, 5232 Villigen, Switzerland}\\
		{\large \textsuperscript{$*$}Corresponding author: uangst@ethz.ch}\\
		\vspace*{0.5cm}
	\end{center}
\section*{Abstract} \label{sec:abstract}
Accurate model predictions of corrosion-driven damage in reinforced concrete structures necessitate a comprehensive understanding of the rate of corrosion product formation. Here, we investigate the influence of dissolved \ch{Si} characteristic of cementitious systems on the rate of corrosion product transformation at alkaline pH. Compared to systems aged in the absence of \ch{Si}, small amounts of \ch{Si} retard the  formation rate of the thermodynamically stable corrosion product goethite by a factor of 10. The estimated first order rate constant of transformation $k$ decreases exponentially as a function of the dissolved \ch{Si} concentration and follows the progression $\text{log}_{10}k = \text{log}_{10}k_0 - 14.65\times[\ch{Si}]^{0.28}$. Findings further suggest that the observed retardation is primarily due to the formation of a mobile aqueous \ch{Fe}-\ch{Si} complex. The concentration of Si in cementitious systems has a crucial influence, and additional research is required to fully incorporate this factor into reactive transport models, ultimately essential for accurate service life predictions.
\end{titlepage}
\section{Introduction} \label{sec:introduction}
Corrosion of steel in concrete is widely recognised as one of the main causes for the structural degradation of reinforced concrete structures\cite{RN89}.
Apart from the corrosion process itself and the reduction in the reinforcement cross-sectional area, the dissolution of iron at the steel-concrete interface may lead to the precipitation of corrosion products in the concrete pore network. Over time, the formation of these iron (hydr)oxide phases can generate internal stresses leading to cracks, spalling of the concrete cover, ingress of chlorides, carbonates and moisture\cite{hongwong_corrosion_products}, thus rendering the degradation process self-sustaining. Within this vicious cycle, both the concrete pore solution chemistry as well as the type of corrosion products formed play a crucial role. \\

Over the course of the last decades, the formation of iron (hydr)oxide phases in aqueous electrolytes has been researched extensively in the fields of soil \cite{schwertmann_ph_ferri,schwertmann_thalman_Si_lepi} and environmental science\cite{twoline_gt_transformation,das_arsenate}. Whilst the majority of studies published focused on their stability at mildly acidic to mildly alkaline pH, selected publications investigated the transformation of corrosion products in mill raffinates\cite{RN229}, air pollution control residues \cite{sorensen_air_pollution_control_residue} and other highly alkaline systems  \cite{hansen_hanford}. In comparison, comparatively little research has been devoted to the formation of corrosion products in highly alkaline environments and in the presence of other solutes commonly encountered in cementitious systems. In an attempt to forecast the corrosion performance of traditional, highly alkaline binders including Portland cement and facilitate the development and usage of alternative, environmentally benign cementitious materials, a more thorough scientific understanding about the formation of corrosion products in aqueous electrolytes characteristic to these systems must be established \cite{angst_achilles}. 

\subsection{Some corrosion products are more expansive than others}
Depending on the crystal structure and the bulk chemical composition, some iron (hydr)oxides have significantly larger specific molar volumes than others. Polymorphs of the \ch{FeOOH} type generally feature a small specific molar volume, ranging from 20.9 \si{\cubic\centi\meter\per\mole} for goethite ($\upalpha-\ch{FeOOH(s)}$) to approximately 25.5 \si{\cubic\centi\meter\per\mole} for akagan\'{e}ite ($\upbeta-\ch{FeOOH(s)}$) \cite{majzlan2003thermodynamics,post1991crystal}.
Amorphous iron hydroxides including 2- and 6-line ferrihydrite ($2,6\text{l}-\ch{Fe(OH)3(s)}$) feature a $V_0$ of up to 34.0 \si{\cubic\centi\meter\per\mole}\cite{dilnesa2011iron} and are thus almost 5 times as voluminous as crystalline \ch{Fe^0} ($V_0=7.10$ \si{\cubic\centi\meter\per\mole})\cite{robie1995thermodynamic}. From a concrete durability viewpoint, the formation of either of these comparatively small \ch{FeOOH} polymorphs is favourable, as the magnitude of internal stresses generated is expected to be lower.

\subsection{The influence of the pore solution chemistry}
The rate and mechanism of corrosion product formation and transformation is heavily dependent on both the pH and the presence of pore solution constituents other than aqueous \ch{Fe}. 2-line ferrihydrite, the most soluble iron hydroxide often preceding the formation of more crystalline iron phases\cite{schwertmann2008iron}, has been observed to exclusively transform to goethite ($\upalpha-\ch{FeOOH(s)}$) at pH > 13\cite{furcas_transformation_est}, i.e. characteristic to the pore solution of the majority of  uncarbonated Portland cement based binders\cite{leemann_loth_2008,leemann_2011,lesaout_2013}. At lower pH values (pH = 9 to 11), as typical for the pore solution of carbonated concrete \cite{RN295}, MgO-based cements \cite{bernard_MgO} and MSH \cite{bernard_MSH}, 2-line ferrihydrite transforms into mixtures of hematite ($\upalpha-\ch{Fe2O3(s)}$) and goethite\cite{twoline_gt_transformation}. As the overall process is rate-limited by the dissolution of 2-line ferrihydrite, transformation rates are greatly accelerated the higher the degree of alkalinity \cite{furcas_speciation_controls_preprint}. \\

Other ionic species including sulphates, carbonates and silicon have been reported to retard phase transformation or completely alter the type(s) of iron (hydr)oxide(s) stabilised. Amongst all ionic constituents commonly encountered in the pore solution of cementitious systems, silicon arguably plays the largest role. Small amounts of \ch{Si} in the orders of $1\times10^{-4}$ M are characteristic for the pore solution of a diverse landscape of cementitious systems including CEM I\cite{leemann_loth_2008,leemann_2011}, CEM I with additions of calcium sulfoaluminates\cite{lesaout_2013}, fly-ash\cite{scholer_2017,deschner2013effect,RN287}, blast-furnace slag\cite{RN287}, silica fume\cite{RN287,bach_SF,chappex_thesis}, metakaolin\cite{chappex_thesis} or CEMIII/B with nanosilica\cite{lothenbach2012hydration}. At pH $\sim$ 13, \ch{Si} retards the dissolution of 2-line ferrihydrite and readily adsorbs onto goethite \cite{schwertmann2008iron,RN280}. \citeauthor{RN294} \cite{RN294} show that the Fe(II)-catalysed transformation of 2-line ferrihydrite and lepidocrocite ($\upgamma-\ch{FeOOH(s)}$) to goethite and magnetite ($\upalpha-\ch{Fe3O4(s)}$) is inhibited in the presence of silicate at neutral pH. Other studies evidence the formation of aqueous \ch{Fe}-\ch{Si} complexes\cite{Pokrovski_iron_silica_EXAFS}, that could influence total aqueous iron concentration. \\

This paper investigates the effect of dissolved silicon on the formation of corrosion products at alkaline pH $\geq 13$. A combination of X-ray diffraction (XRD) and X-ray absorption spectroscopy (XAS) measurements is used to determine the type and formation rate of iron (hydr)oxide phases aged in the presence of 0.1 and 0.5 mM \ch{Na2SiO3}. Aqueous \ch{Fe} and \ch{Si} concentration measurements determined via inductively coupled plasma optical emission spectroscopy (ICP-OES) provide further insights into the likely mechanism of corrosion product transformation, the effect of silicon on the dissolution and re-precipitation of solid phases involved and the formation of aqueous \ch{Fe}-\ch{Si} complexes. In combination with the results of a previous study investigating the rate of corrosion product transformation in the absence of silicon, we derive a semi-empirical relationship that quantifies the estimated first-order rate constant of transformation $k$ as a function of the aqueous \ch{Si} concentration. Findings can be used to calculate the rate of iron (hydr)oxide formation across a broad range of cementitious systems. This study is intended to serve as a base-case for future investigations into the formation of corrosion products in the presence of sulfates, carbonates and other prominent cementitious pore solution constituents. 
\section{Methods} \label{sec:methods}
\subsection{Preparation of supersaturated stock solutions}
The sample preparation has been described in detail in \citeauthor{furcas_transformation_est} \cite{furcas_transformation_est}. Briefly, 1 M of $\ch{FeCl3}\cdot6\ch{H2O(cr)}$ has been dissolved in 2 wt. \% \ch{HNO3}. 5 \si{\milli\liter} of the resultant acidic stock solution was mixed with 245 \si{\milli\liter} of 4 basic stock solutions containing 0.104 M (pH = 13) and 1.024 M (pH = 14) \ch{NaOH} as well as 0.100 and 0.500 mM of \ch{Na2SiO3}, respectively. Stock solutions were stirred rigorously and aged in fresh polyethylene containers at ambient temperature for up to 30 days. Solid phases were extracted by centrifugation at 10000 rpm for 15 min and subsequent freeze drying and stored as dry powders. Liquid phase aliquots were taken by extracting approx. 1 \si{\gram} from the aged stock solution with a syringe. Aliquots were filtered using 0.20 \si{\micro\meter} nylon filters (Semadeni AG, Ostermundigen, Switzerland) and mixed with 2 wt. \% \ch{HNO3} (EMSURE) purchased from Merck Group (Merck KGaA, Darmstadt, Germany), at a ratio of 1:10 for early and 1:8 for late equilibration times to prevent further precipitation.
\subsection{Iron hydroxide reference standards}
A total of 3 reference standards are considered in this study, 2-line ferrihydrite ($2\text{l}-\ch{Fe(OH)3(s)}$), goethite ($\alpha-\ch{FeOOH(s)}$) and lepidocrocite ($\gamma-\ch{FeOOH(s)}$). They were selected based on a previous investigation into the formation of corrosion products at alkaline pH\cite{furcas_transformation_est}. 
Both 2-line ferrihydrite and lepidocrocite were synthesised following the experimental procedure of \citeauthor{schwertmann2008iron} \cite{schwertmann2008iron}, goethite was purchased from Thermo Fisher Scientific, Waltham, MA, USA. The identity of all standards was confirmed by XRD\cite{furcas_transformation_est}. Their extended X-ray absorption fine structure (EXAFS) spectra were recorded in a previous study \cite{furcas_transformation_est}.
\subsection{Inductively coupled plasma optical emission spectroscopy}
The composition of the solution was measured using the Agilent 5110 ICP-OES instrument (Agilent Technologies Inc., Santa Clara, CA, USA). Iron and silica concentrations were determined via linear interpolation on a set of 8 discrete reference concentrations, ranging from 0.01 to 50 ppm for \ch{Fe} and 0.01 to 50 ppm for \ch{Si}. The limit of quantitation (LOQ) was determined for each analyte emission line, \ch{Fe} 234.350, 238.204, 239.563, 259.940 \si{\nano\meter} and \ch{Si} 250.690, 251.611, 288.158 \si{\nano\meter}, according to the reccomendations of \citeauthor{RN147} \cite{RN147} (Supporting information, Tab. \ref{tab:ICP_LOD_LOQ}). At 0.163 and 1.105 \si{\micro\mole\per\liter}, \ch{Fe} 259.940 \si{\nano\meter} and \ch{Si} 250.690 \si{\nano\meter} feature the highest LOQ of all analyte spectral lines recorded and will thus be considered the global limits of quantitation. The detailed account of all reference standard compositions and calibration lines is presented in Supporting Information, Tab. \ref{tab:ICP_Standards_noSi}, \ref{tab:ICP_Standards_Si}. and Fig. \ref{fig:ICP_calibration_Fe_8}, \ref{fig:ICP_calibration_Fe} and \ref{fig:ICP_calibration_Si}.
For more details, the reader is referred to \citeauthor{furcas_transformation_est} \cite{furcas_transformation_est}.
\subsection{X-ray diffraction}
Powder X-ray diffraction patterns were recorded from $4^\circ$ to $80^\circ$ in steps of 0.02 $2\uptheta$, using the Bruker D8 Advance. The diffractometer uses a coupled $\uptheta-2\uptheta$ configuration, Co K$\upalpha$ ratiation ($\lambda = 1.7902$ \si{\angstrom}) and a LynxEye XE-T detector. Diffraction patterns of reference components and solids extracted from supersaturated stock solution were analysed with the open source X-ray diffraction and Rietveld and refinement programme Profex \cite{doebelin_profex}. The identity of the reference standards was confirmed by comparison to the reference powder diffraction spectra of 2-line ferrihydrite \cite{structure_ferri}, goethite \cite{brown_XRD,harrison_goethite} and lepidocrocite (PDF entry 00-044-1415), as detailed in \cite{furcas_transformation_est}.   
\subsection{X-ray absorption spectroscopy}
All synchrotron-based investigations were carried out at the PHOENIX (Photons for the Exploration
of Nature by Imaging and XAFS) beamline at the Swiss Light Source (SLS), Paul Scherrer Institute (PSI), Villigen, Switzerland. The beamline specifications have been described elsewhere\cite{henzer_calcium_carbonate}. Iron hydroxide powders were uniformly applied to conductive carbon tape on a roughened-up copper sample holder and measured at ambient temperature under vacuum at approx. $10^{-6}$ \si{\bar}. The Total Fluorescent Yield (TFY) of the sample was measured using a four-element vortex detector. The Total Electron Yield (TEY) of the sample was measured as the incident photon flux on a Ni-coated polyester foil, located in a different vacuum chamber at approx. $5\cdot10^{-8}$ \si{\bar}. Iron hydroxide powder samples were measured $3-5$ times, the reference standards 8 times.

\noindent Fe K-edge spectra including the X-ray absorption near edge structure (XANES) and the extended X-ray absorption fine structure (EXAFS) were recorded from 7041 to 7858 eV, i.e. from 70 eV below to approximately 750 eV above the absorption edge. The recorded data was analysed and normalised using the IFFEFIT (ATHENA) software package\cite{RN156,RN155}. The attenuated TFY signal was corrected for self-absorption by adjusting it to the TEY using the software in-built Fluo-$\upmu(E)$ algorithm, providing the molecular formula of various samples analysed. Radial distribution functions were generated from $k^3$-weighted EXAFS spectra by performing a Fourier transform 
over 1.5 to 10.0 \si{\per\angstrom}, using the Kaiser-Bessel window function. The relative amount of different iron hydroxides in the sample was determined by linear combination fitting (LCF) in the XANES region extending from 20 eV below to 40 eV above the Fe K-edge. The weights of all reference standards, 2-line ferrihydrite, lepidocrocite and goethite, were constrained to lie between the values of 0 and 1 and forced to add up to 1.

\section{Results and discussion} \label{sec:results}
\subsection{Aqueous concentration measurements}
\begin{figure}[!ht]
	\centering
	%\hspace{-25pt}
	\begin{subfigure}[b]{0.5\textwidth}
		\centering
		\includegraphics[scale=0.77]{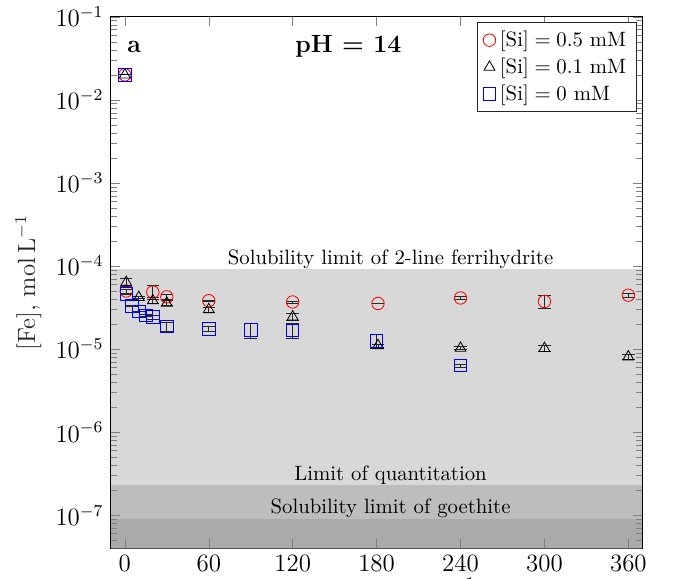}
	\end{subfigure}%
	%\vskip\baselineskip
	\begin{subfigure}[b]{0.5\textwidth}
		\hspace{0.2cm}
		\centering
		\includegraphics[scale=0.77]{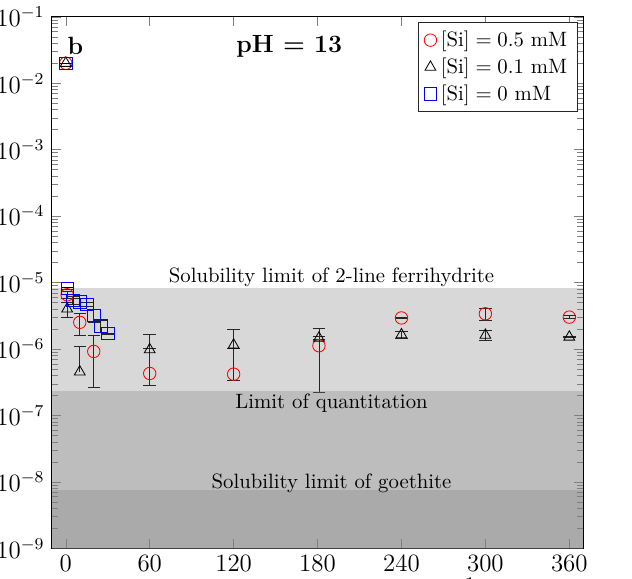}
	\end{subfigure}
	\vskip\baselineskip
	\begin{subfigure}[b]{0.5\textwidth}
		\centering
		\includegraphics[scale=0.77]{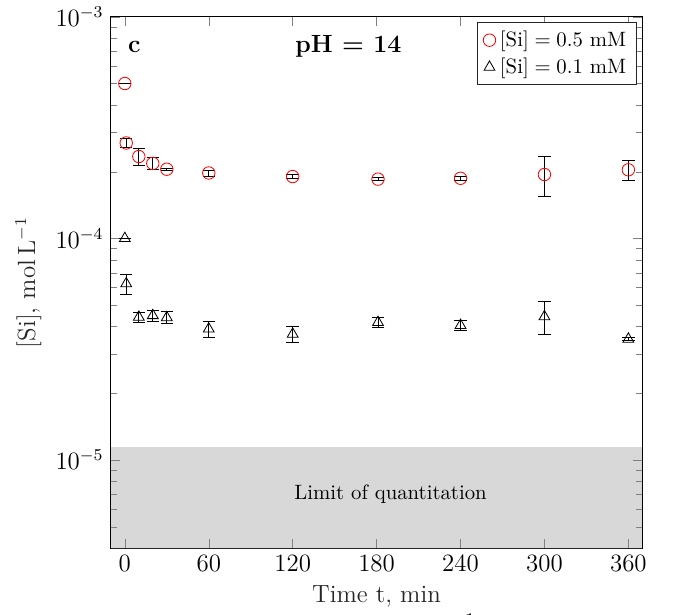}
	\end{subfigure}%
	%\vskip\baselineskip
	\begin{subfigure}[b]{0.5\textwidth}
		\hspace{0.2cm}
		\centering
		\includegraphics[scale=0.77]{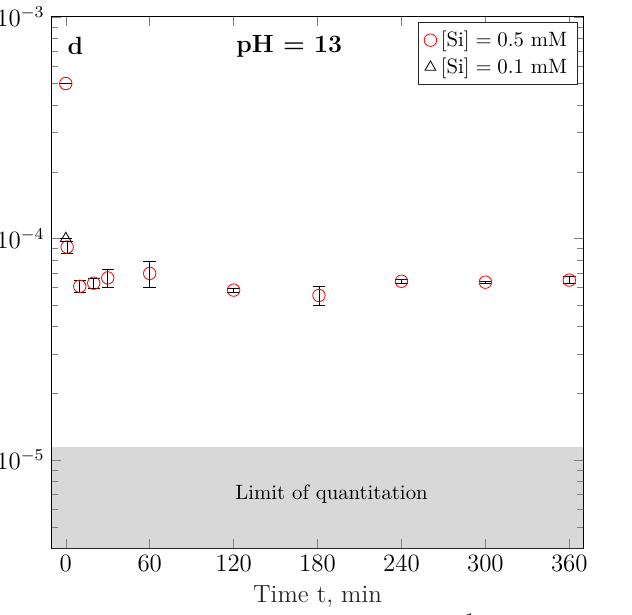}
	\end{subfigure}
	\caption{Total aqueous concentration of iron (Fig. \ref{fig:ICP}a and Fig. \ref{fig:ICP}b) and silicon (Fig. \ref{fig:ICP}c and Fig. \ref{fig:ICP}d) at pH = 14 (Fig. \ref{fig:ICP}a and Fig. \ref{fig:ICP}c) and pH = 13 (Fig. \ref{fig:ICP}b and Fig. \ref{fig:ICP}d) over time. Concentrations are measured by ICP-OES at the emission lines \ch{Fe} 259.940 \si{\nano\meter} and \ch{Si} 250.690 \si{\nano\meter}. The error bars displayed represent the standard deviation of the determined elemental concentrations in three independent experimental runs. Note that for the measurements of \ch{Si}-free solutions, the level of quantitation amounts to $1\times10^{-6}$ \si{\mole\per\liter} due to the increased dilution with 2 wt. \% nitric acid\cite{furcas_transformation_est}.}
	\label{fig:ICP}
\end{figure}

The aqueous iron and silicon concentrations
measurements in equilibrium with various solid iron hydroxide phases suggest the transformation of highly soluble 2-line ferrihydrite towards a more crystalline Fe-bearing phase. Fig. \ref{fig:ICP} show the decrease of $[\ch{Fe}]$ and $[\ch{Si}]$ as a function equilibration time. At pH = 14 (Fig. \ref{fig:ICP}a and Fig. \ref{fig:ICP}c), the initial reduction in the aqueous iron concentration observed in silicon-free solutions prevails across all silicon concentrations tested. Levelling off at approximately $5\times10^{-5}$ M after 1 hour, the aqueous iron concentration in the presence of 0.5 mM \ch{Na2SiO3} is 5 times higher than that measured in the absence of silicon. The iron concentration in solutions containing 0.1 mM \ch{Na2SiO3}  decreases faster than that measured in the presence of 0.5 mM \ch{Na2SiO3} and slower compared to the silicon-free samples.
After 3 hours, the aqueous iron concentration in the presence of both 0 mM and 0.1 mM \ch{Na2SiO3} plateaus at $\sim1\times10^{-5}$ M, i.e. approximately 100 times higher than the calculated solubility of goethite at pH 14. All measured $[\ch{Fe}]$, within 1 minute of equilibration time and thereafter, were found to be below the solubility limit of 2-line ferrihydrite and approach the aqueous iron concentration in equilibrium with goethite at rates inversely correlated to the initial concentration of \ch{Na2SiO3}. The aqueous silicon concentration on the other hand rapidly drops to 40 \% of its initial value, irrespective of the dissolved amounts of \ch{Na2SiO3}.
At pH = 13 (Fig. \ref{fig:ICP}b and Fig. \ref{fig:ICP}d), the progression of $[\ch{Si}]$ rapidly reduces to $\sim6\times10^{-5}$ M, i.e. approximately 10 \% of the initial concentration. At the same time, aqueous iron concentrations are significantly lower than those measured at pH = 14. Even though the solubility limit of 2-line ferrihydrite decreases by one order of magnitude from pH 14 to 13 (compare Figure \ref{fig:ICP}a and  \ref{fig:ICP}b), the same initial reduction in the aqueous iron concentration within 1 minute of equilibration time is observed. No clear correlation can be established between the decrease of  $[\ch{Fe}]$ and the dissolved aqueous silicon concentration. 
Within the standard deviation of various $[\ch{Fe}]$ in the presence of 0.1 mM and 0.5 mM \ch{Na2SiO3}, it appears that the aqueous iron concentration initially decreases below the measured concentrations in the absence of silicon, but then increases above it at longer equilibration times, striving towards a consistent solubility of $2\times10^{-6}$ after 2 hours. The $[\ch{Fe}]$ in the presence of \ch{Na2SiO3} is in the order of $3\times10^{-7}$ M within the first 2 hours of equilibration, just above the LOQ of the ICP device, and thus features significantly higher standard errors than the measurements at pH = 14. Note that the aqueous silicon concentration measured from solutions containing 0.1 mM \ch{Na2SiO3} at pH = 13 are below the respective LOQ across all analyte emission lines.  
\newpage
\clearpage
\subsection{Mineral phase identification}
\begin{figure}[!ht]
	\centering
	%\hspace{-25pt}
	\begin{subfigure}[b]{\textwidth}
		\centering
		\includegraphics[scale=0.85]{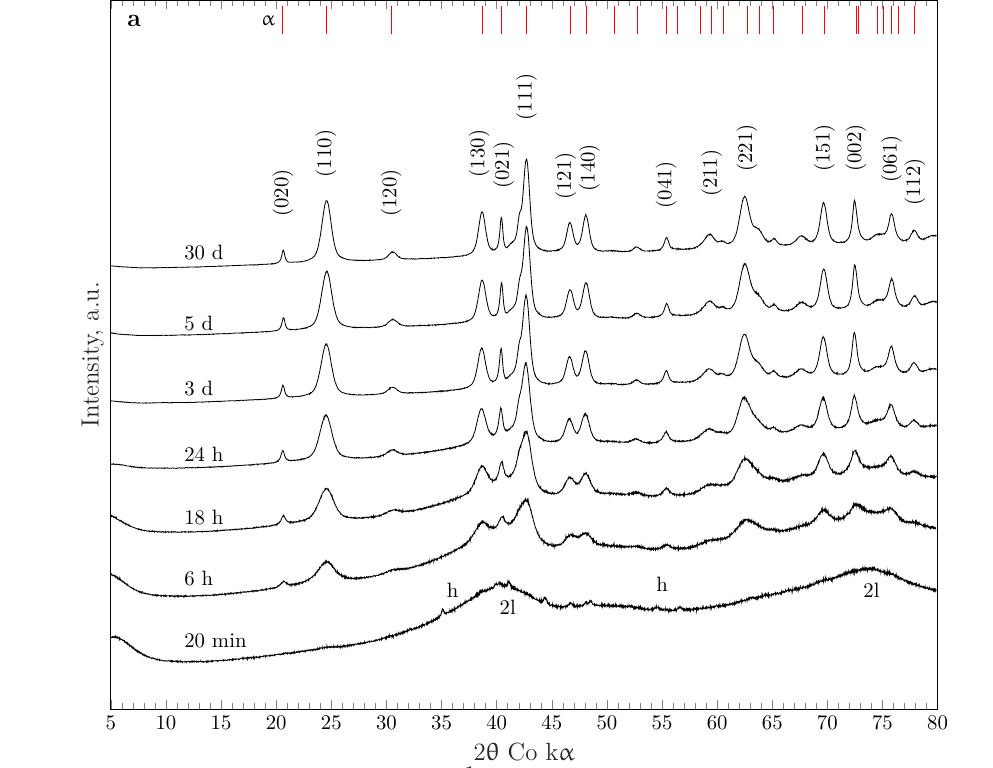}
	\end{subfigure}%
	\vskip\baselineskip
	\begin{subfigure}[b]{\textwidth}
		%\hspace{0.2cm}
		\centering
		\includegraphics[scale=0.85]{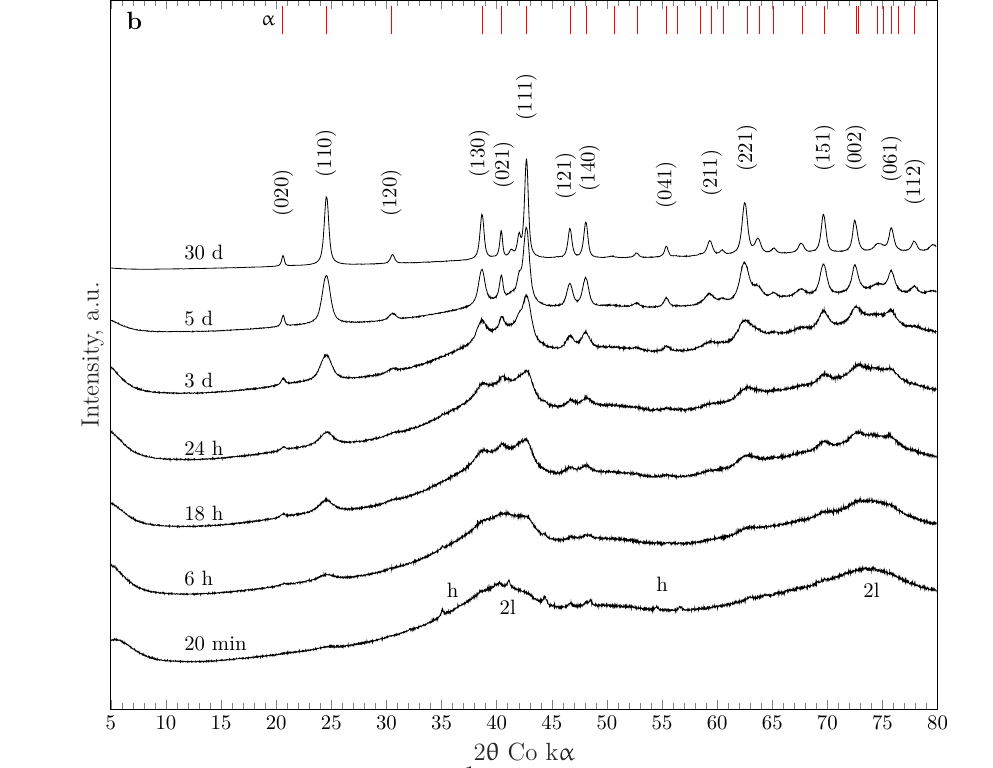}
	\end{subfigure}
	\caption{XRD patterns of iron hydroxide phases extracted from stock solutions containing 20.0 mM $\ch{FeCl3}\cdot6\ch{H2O(cr)}$ and 0.1 mM \ch{Na2SiO3} (Fig. \ref{fig:XRD_pH14}a) and 0.5 mM \ch{Na2SiO3} (Fig. \ref{fig:XRD_pH14}b) at pH = 14. Main peak positions of goethite ($\alpha-\ch{FeOOH(s)}$), 2-line ferrihydrite ($2\text{l}-\ch{Fe(OH)3(s)}$) and halite (\ch{NaCl(s)}) are indicated with a red vertical bar (\color{red}$\lvert$\color{black}), 2l and an h.}
	\label{fig:XRD_pH14}
\end{figure}

\begin{figure}[!ht]
	\centering
	\begin{subfigure}[b]{0.4\textwidth}
		\hspace{-4cm}
		\includegraphics[scale=0.85]{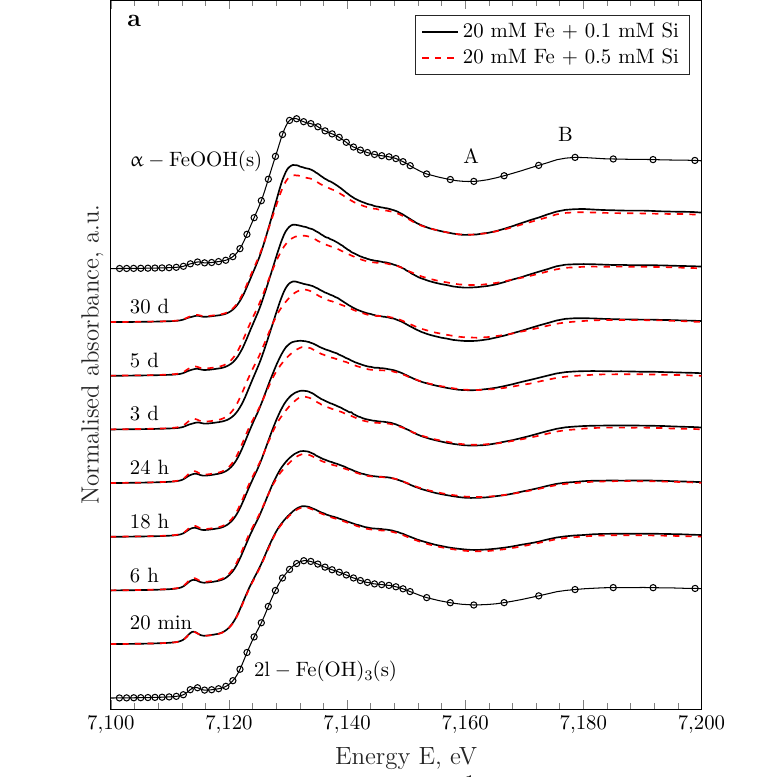}
	\end{subfigure}%
	%\vskip\baselineskip
	\begin{subfigure}[b]{0.2\textwidth}
		\hspace{-1cm}
		\includegraphics[scale=0.85]{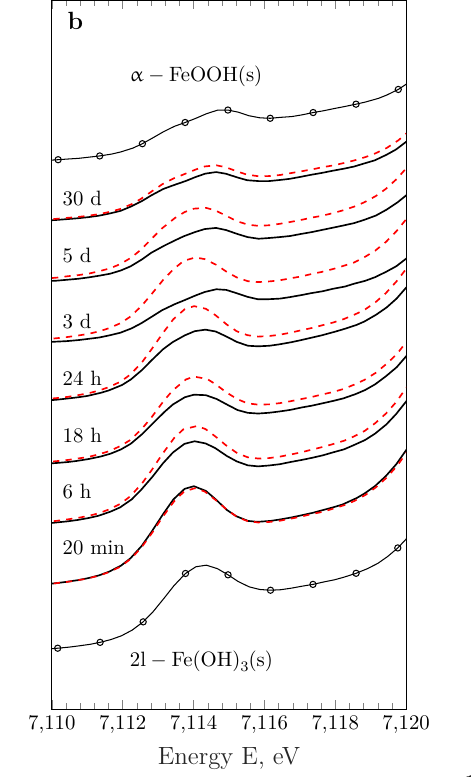}
	\end{subfigure}
	\caption{Normalised Fe K-edge XANES spectra of iron hydroxide powders  extracted from supersaturated stock solutions at pH = 14 containing 20.0 mM $\ch{FeCl3}\cdot6\ch{H2O(cr)}$ and 0.1 mM \ch{Na2SiO3} (black continuous lines)
		and 0.5 mM \ch{Na2SiO3} (red dashed lines), together with the reference standards spectra of 2-line ferrihydrite ($2\text{l}-\ch{Fe(OH)3(s)}$) and goethite ($\upalpha-\ch{FeOOH(s)}$). Fig. \ref{fig:XANES}a highlights the white line intensity as well as the first destructive (A) and constructive (B) interference after the absorption edge, Fig. \ref{fig:XANES}b shows a zoomed-in section of the pre-edge feature.}
	\label{fig:XANES}
\end{figure}

\begin{figure}[!ht]
	\centering
	\begin{subfigure}[b]{0.4\textwidth}
		\hspace{-4cm}
		\includegraphics[scale=0.85]{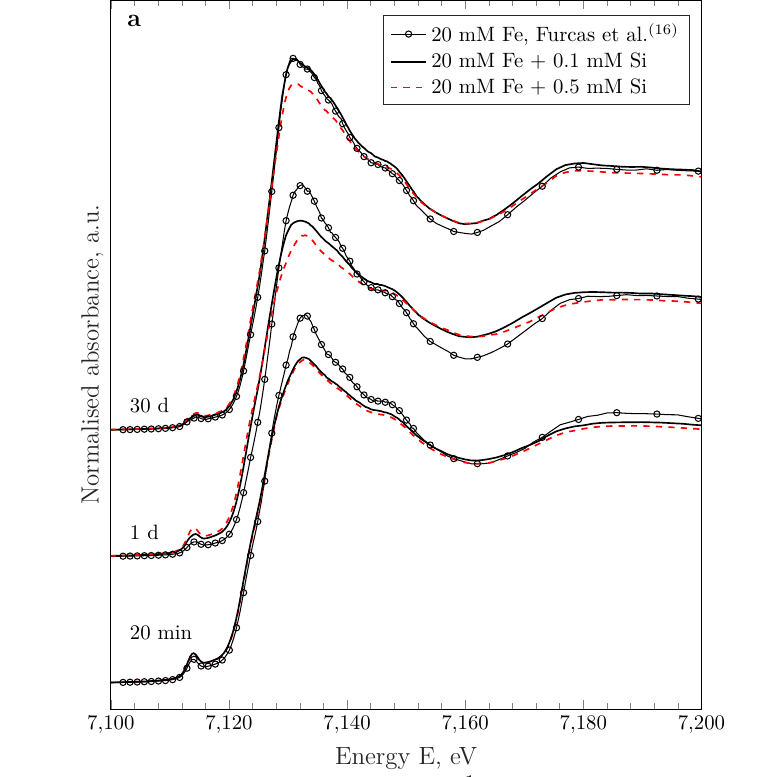}
	\end{subfigure}%
	%\vskip\baselineskip
	\begin{subfigure}[b]{0.2\textwidth}
		\hspace{-1cm}
		\includegraphics[scale=0.85]{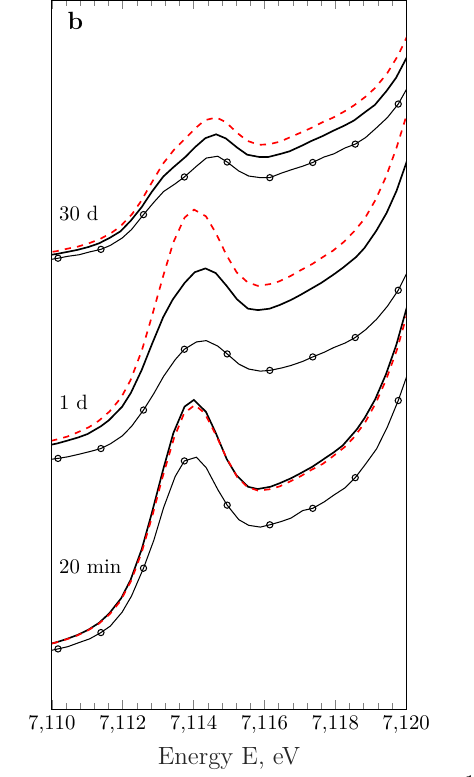}
	\end{subfigure}
	\caption{Normalised Fe K-edge XANES spectra of iron hydroxide powders extracted from solutions containing 20.0 mM $\ch{FeCl3}\cdot6\ch{H2O(cr)}$ and no additional silicon\cite{furcas_transformation_est},  0.1 mM \ch{Na2SiO3} (black continuous lines) and 0.5 mM \ch{Na2SiO3} (red dashed lines).}
	\label{fig:XANES_comparison}
\end{figure}

\begin{figure}[!ht]
	\centering
	\begin{subfigure}[b]{\textwidth}
		\includegraphics[scale=0.85]{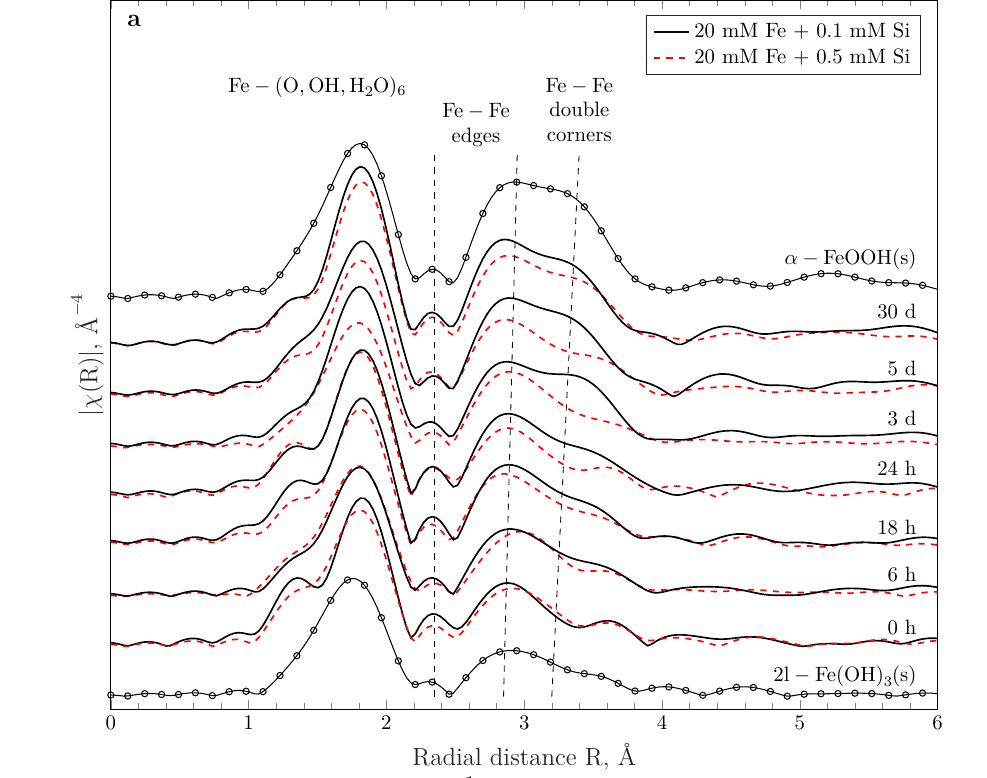}
	\end{subfigure}%
	\vskip\baselineskip
	\begin{subfigure}[b]{\textwidth}
		\includegraphics[scale=0.85]{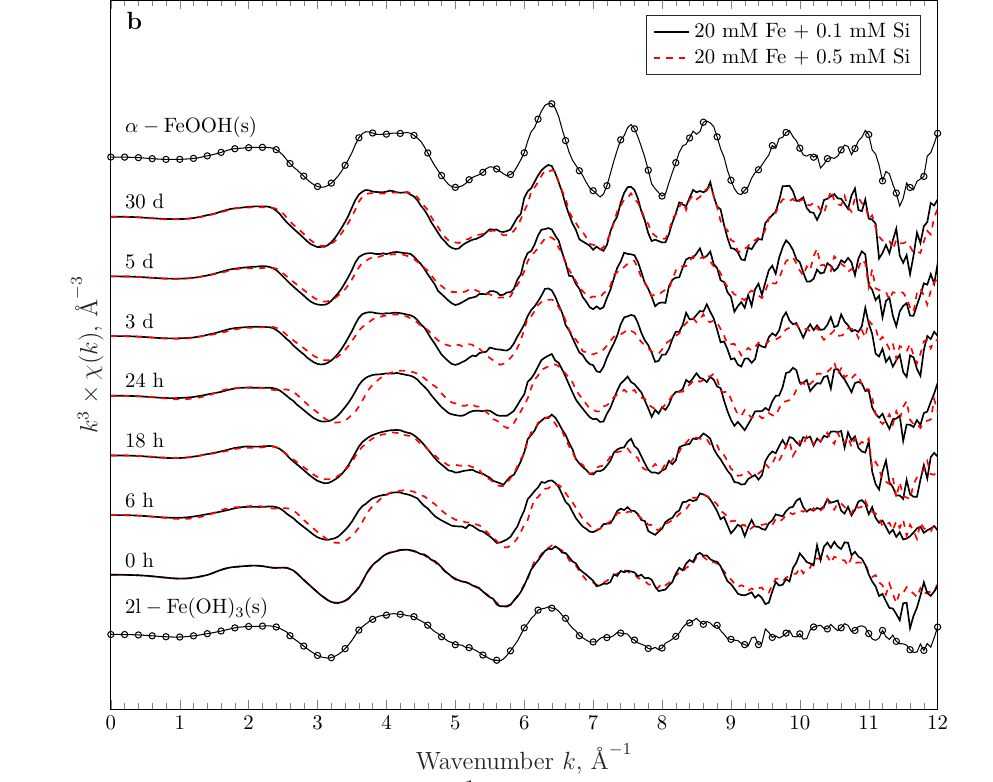}
	\end{subfigure}
	\caption{Fourier transforms of $k^3$-weighted Fe K-edge EXAFS spectra of iron hydroxide powders extracted from supersaturated stock solutions at pH = 14 (Fig. \ref{fig:EXAFS_R_space}a), together with the $k^3$-weighted EXAFS spectra (Fig. \ref{fig:EXAFS_R_space}b). Solid black and dashed red lines correspond to solutions containing 20.0 mM $\ch{FeCl3}\cdot6\ch{H2O(cr)}$ and 0.1 mM \ch{Na2SiO3} and  0.5 mM \ch{Na2SiO3}, respectively. The Fourier transform was performed using the Kaiser-Bessel window function over a $k$-space ranging from 1.5 to 10.0 \si{\per\angstrom}.}
	\label{fig:EXAFS_R_space}
\end{figure}

\begin{figure}[!ht]
	\centering
	\includegraphics[scale=0.90]{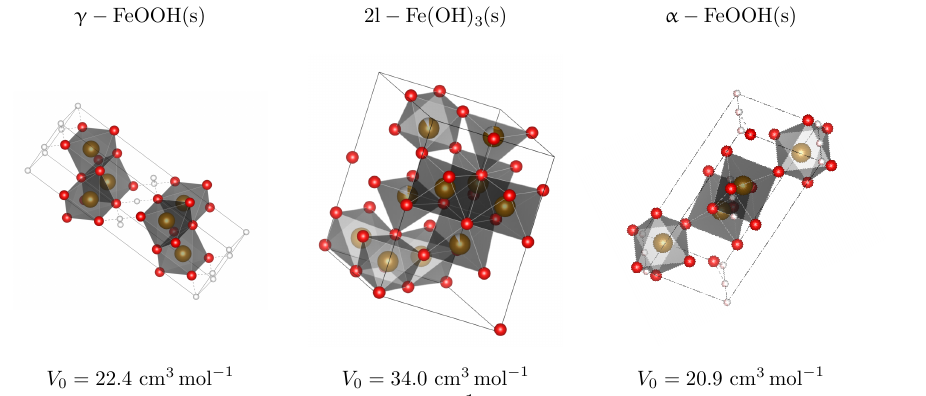}
	\caption{The structural models of iron hydroxides lepidocrocite ($\upgamma-\ch{FeOOH(s)}$), 2-line ferrihydrite ($2\text{l}-\ch{Fe(OH)3(s)}$) and goethite ($\upalpha-\ch{FeOOH(s)}$), together with their specific molar volumes in \si{\cubic\centi\meter\per\mole}. The central iron atoms are depicted in orange, oxygen in red and hydrogen atoms in white. The structural models are generated from the .cif files of their respective reference standards, plotted in Vesta.} 
	\label{fig:XAS_crystal}
\end{figure}

\begin{figure}[!ht]
	\centering
	\includegraphics[scale=0.85]{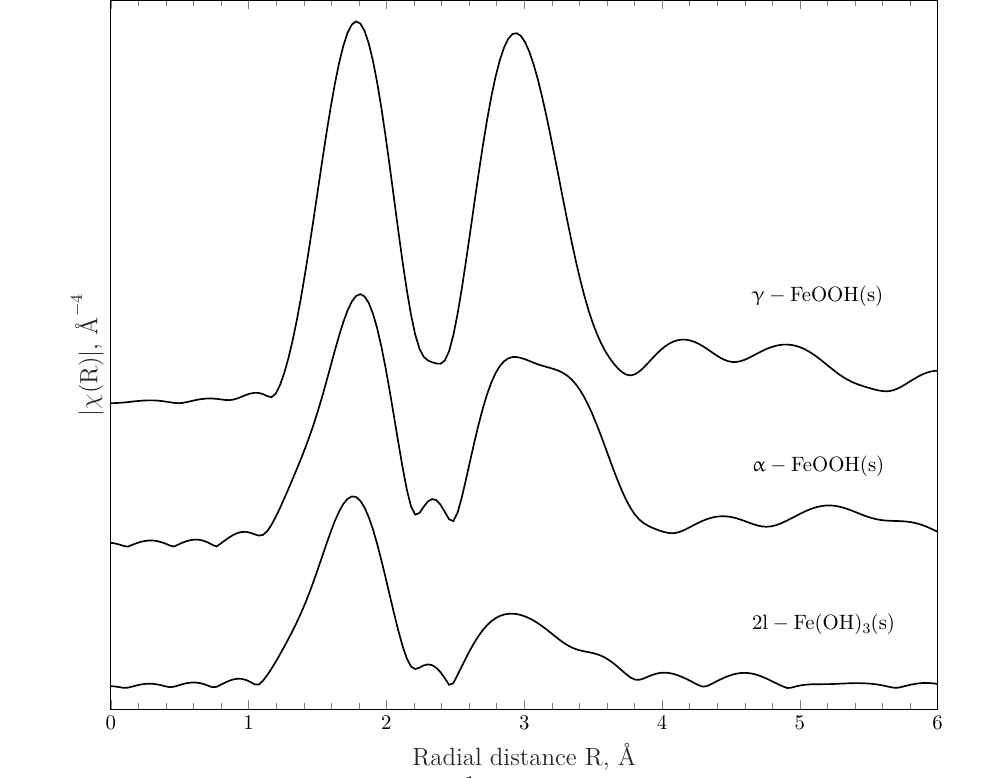}
	\caption{Fe K-edge $k^3$-weighted RDF of solid iron hydroxide reference standards 2-line ferrihydrite ($2\text{l}-\ch{Fe(OH)3(s)}$), lepidocrocite ($\gamma-\ch{FeOOH(s)}$) and goethite ($\alpha-\ch{FeOOH(s)}$), together with the graphical representation of their unit-cell crystal structure. Octahedra making up the unit cell of lepidocrocite are predominantly joined at their faces and consequentially feature a small $\ch{Fe}-\ch{Fe}$ double corner contribution. The RDF of goethite entails a close to equal contribution of $\ch{Fe}-\ch{Fe}$ edges and double corners, as evident from the topology of its unit cell\cite{manceau_combes_Fe_topology}. The ferric cation in 2-line ferrihydrite on the other hand is coordinated both tetrahedrally and octahedrally\cite{structure_ferri}. The corresponding RDF features two discernable, low-intensity contributions from $\ch{Fe}-\ch{Fe}$ edges as well as double corners.} 
	\label{fig:XAS_standards_crystal}
\end{figure}

The obtained X-ray diffraction patterns and absorption spectra independently corroborate the formation of goethite from 2-line ferrihydrite at the investigated physiochemical conditions.
As illustrated in Fig. \ref{fig:XRD_pH14}, dried solids extracted from supersaturated iron stock solution at pH = 14 are completely amorphous to X-ray diffraction at early equilibration times. Both diffraction patterns at 20 minutes feature two broad peaks at $\sim40^\circ$ and $\sim74^\circ$ $2\uptheta$, characteristic to amorphous 2-line ferrihydrite as well as negligible amounts of halite (\ch{NaCl(s)}). In the presence of 0.1 mM \ch{Na2SiO3} (Fig. \ref{fig:XRD_pH14}a), all major peaks of goethite $\upalpha-\ch{FeOOH(s)}$ emerge from the initial diffraction pattern within one day, similar to the observations in the absence of \ch{Si}\cite{furcas_transformation_est}. At elevated concentrations of 0.5 mM \ch{Na2SiO3} (Fig. \ref{fig:XRD_pH14}b), only the most prominent lattice plane reflections are discernible from the amorphous diffraction pattern of 2-line ferrihydrite after 1 day. Peaks after 30 days are generally sharper than those measured in the 0.1 mM \ch{Na2SiO3} sample or in the absence of silicon (Supporting information, Fig. \ref{fig:XRD_peaks}) at the same point in time. The observed increase in peak sharpness can be attributed to the adsorption of \ch{Si} onto the terminal goethite crystal faces, leading to preferential growth on the other crystal faces\cite{schwertmann2008iron}.
\newpage
\clearpage
Experimentally determined Fe K-edge XANES spectra  (Fig. \ref{fig:XANES}) exhibit a pre-edge feature at approximately $7113.5\pm0.5$ eV, resulting from quadrupolar 1s $\rightarrow$ 3d electronic transitions in the absorber iron ion, characteristic to Fe(III) compounds \cite{wilke_XAS,apted_iron_structure}. Over the course of the experiment, the integrated peak intensity of the pre-edge feature reduces, whilst the white line intensity as well as the magnitude of the first destructive (A) and constructive (B) interference after the rising edge increases, indicating a transition towards a more centrosymetric, octrahedral coordination environment \cite{waychunas_XANES,guda_understanding_XAS}.
Here, the white line and pre-edge feature intensities of spectra corresponding to an initial silicon concentration of 0.1 mM are in between those aged in the presence of 0.5 mM \ch{Na2SiO3} and in the absence of silicon (Fig. \ref{fig:XANES_comparison}). XANES spectra of iron hydroxide powders extracted from supersaturated stock solution after 30 days closely resemble the spectrum of the goethite reference standard.
The observed transformation is also evidenced from the radial distribution functions (RDF) of iron hydroxide powders extracted at different equilibration times. As illustrated in Fig. \ref{fig:EXAFS_R_space}, the peaks in the Fourier transformed $k^3$-weighted EXAFS spectra constitute 3 distinct contributions due to the presence of $\ch{Fe}-\ch{O}$ bonds at distances of $\sim2.0$ \si{\angstrom} \cite{RN182}, $\ch{Fe}-\ch{Fe}$ edges at $\sim2.4$ \si{\angstrom} to $\sim2.9$ \si{\angstrom} and $\ch{Fe}-\ch{Fe}$ double corners at $\sim3.5$ \si{\angstrom}, respectively\cite{Pokrovski_iron_silica_EXAFS,manceau_combes_Fe_topology}. At low equilibration times, the coordination environment of \ch{Fe} resembles that of 2-line ferrihydrite, featuring little to no contribution of the octahedra-linking  $\ch{Fe}-\ch{Fe}$ corners. At successively higher equilibration times, both $\ch{Fe}-\ch{Fe}$ pairs in the two first iron coordination shells become significantly more pronounced. The evolution clearly indicates a change in the local coordination environment of iron and the formation of a corner-sharing $\ch{Fe}-(\ch{O},\ch{OH},\ch{H2O})_6$ octahedral structure.
Moreover, the ratio of the second ($\ch{Fe}-\ch{Fe}$ edge) to the third ($\ch{Fe}-\ch{Fe}$ double corner) peak intensity decreases as a function of equilibration time. At 30 days, the relative peak intensities of both contributions fall short of those measured in the synthetic goethite reference standard. This is presumably due to a lower degree of crystallinity of the goethite phase stabilised at alkaline pH and in the presence of silicon when compared to its synthetic counterpart. The RDF of synthetic lepidocrocite only exhibits an intense $\ch{Fe}-\ch{Fe}$ edge peak and a comparatively weak corner contribution. This feature is unique amongst all common iron (hydr)oxides and arises from the topology of lepidocrocite chains.
\newpage
\clearpage
\newpage
\clearpage
As schematically illustrated in Fig. \ref{fig:XAS_crystal}, one octahedron of lepidocrocite ($\upgamma-\ch{FeOOH(s)}$) shares a total of 6 edges and consequentially only 2 corners with the adjacent ones \cite{manceau_combes_Fe_topology}. 
The respective \ch{Fe}-\ch{Fe} double corner contributions in 2-line ferrihydrite ($2\text{l}-\ch{Fe(OH)3(s)}$) and goethite ($\upalpha-\ch{FeOOH(s)}$), as displayed in Fig. \ref{fig:XAS_standards_crystal} are significantly higher. It is thus likely that the measured iron hydroxide powders do not contain lepidocrocite in quantities sufficiently large to be detected by EXAFS. Comparing the obtained RDF at the same equilibration times, it can be recognised that both the first coordination shell $\ch{Fe}-\ch{O}$ as well as various $\ch{Fe}-\ch{Fe}$ edge and double corner contributions are suppressed in the presence of 0.5 mM, compared to 0.1 mM \ch{Na2SiO3}. Analogous to the hindered  polymerisation of octahedrally coordinated Fe(III) in Si containing solution, as described by  \citeauthor{Pokrovski_iron_silica_EXAFS} \cite{Pokrovski_iron_silica_EXAFS}, the observed peak intensity attenuation strongly suggests the presence of \ch{Si} ligands within the second coordination shell around the ferric ion.
The observed retardation is also in agreement with a number of studies investigating the transformation of solid primary to secondary iron (hydr)oxides and is commonly attributed to the adsorption or incorporation of silicon into the crystal or nucleus of the primary phase \cite{schwertmann_thalman_Si_lepi,RN279,RN280,schwertmann_fischer_1973_ferric}. 
\subsection{The kinetics of corrosion product phase transformation}
\begin{figure}[!ht]
	\centering
	\includegraphics[scale=0.85]{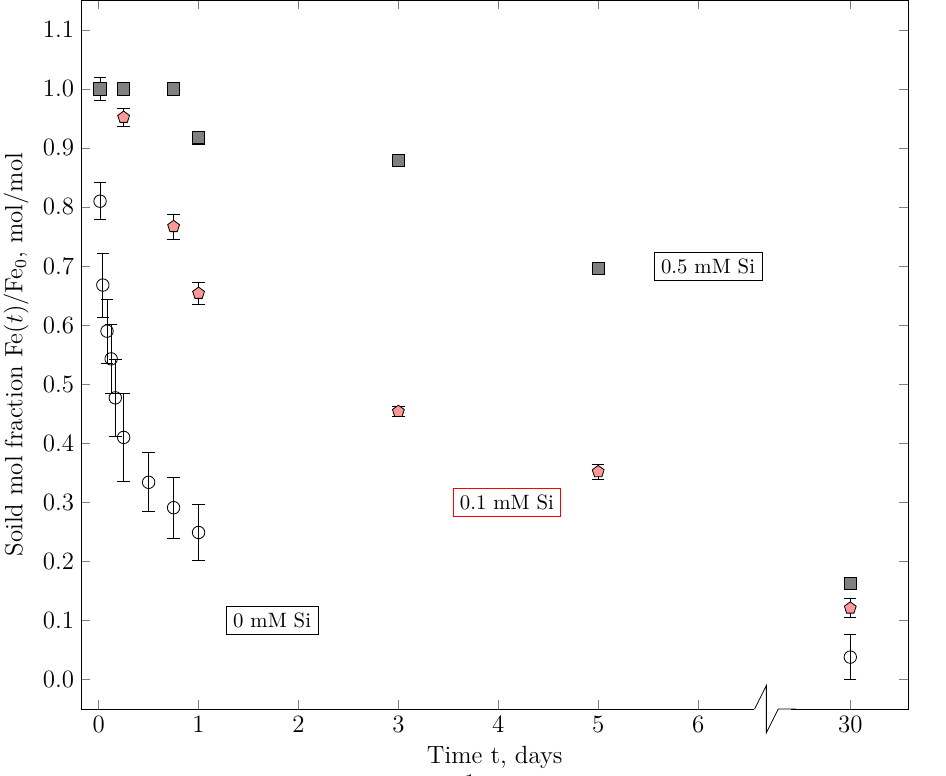}
	\caption{Time dependent molar fractions of 2-line ferrihydrite in the presence of 0, 0.1 and 0.5 mM \ch{Si} at pH = 14. The data for 0 mM \ch{Si} (circular empty markers) is taken from \citeauthor{furcas_transformation_est} \cite{furcas_transformation_est}. Molar fractions at 0.1 mM and 0.5 mM \ch{Si} are obtained from LCF of the XANES spectra displayed in Fig. \ref{fig:XANES_LCF}, using the reference standards of 2-line ferrihydrite and goethite.}
	\label{fig:XANES_LCF_progression}
\end{figure}

\begin{figure}[!ht]
	\centering
	\includegraphics[scale=0.85]{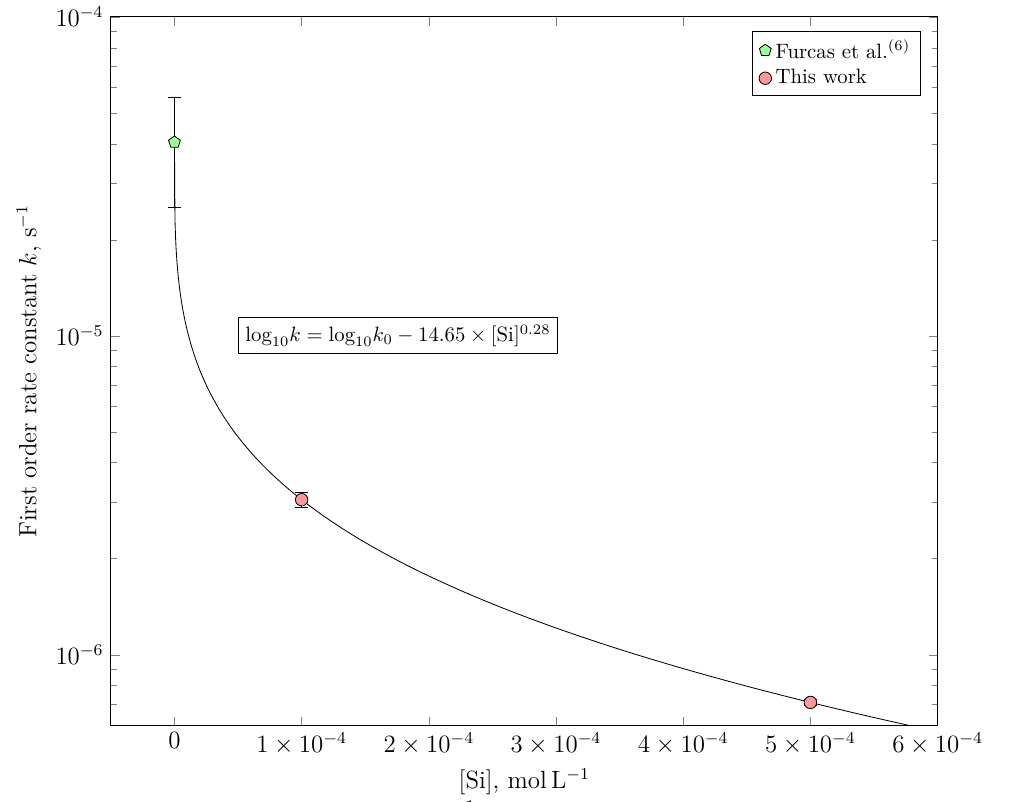}
	\caption{The estimated first order rate constants of corrosion product transformation $k$ in \si{\per\second} as a function of the aqueously dissolved silicon concentration $[\ch{Si}]$ in \si{\mole\per\liter} at pH = 14. Various $k$ were computed by fitting the time dependent molar fraction of 2-line ferrihydrite to the integrated first order rate equation $\ch{Fe}(t) = \ch{Fe}_0\times\text{exp}(-kt)$. The dataset corresponding to the rate of transformation in the absence of silicon was taken from \citeauthor{furcas_transformation_est} \cite{furcas_transformation_est}.}
	\label{fig:rate_constants}
\end{figure}

The rate of phase transformation was calculated by linear combination fitting (LCF) of various XAS spectra of iron hydroxide powders aged in the presence of 0.1 and 0.5 mM \ch{Si} in the near-edge region from 20 eV below to 40 eV above the absorption edge (Supporting information, Fig. \ref{fig:XANES_LCF}). From within the pool of reference standards employed, 2-line ferihydrite, goethite and lepidocrocite, weighted combinations of 2-line ferrihydrite and goethite best reproduce the measured XANES spectra of aged iron hydroxide powders. Including the third standard, lepidocrocite, yielded a less robust fit with comparatively high standard deviations in the predicted solid molar fractions (Supporting information, Fig. \ref{fig:XANES_LCF_progression_lp}). The computed molar fractions of 2-line ferrihydrite as a function of time are displayed in Fig. \ref{fig:XANES_LCF_progression}.  
It is clear that there is a positive correlation between the dissolved silicon concentration and the observed solid fraction of 2-line ferrihydrite. In the absence of silicon, the formation of goethite is rapid, consuming close to 75 \% of 2-line ferrihydrite within 24 hours, as reported in \citeauthor{furcas_transformation_est} \cite{furcas_transformation_est}. In the same timespan, the conversion extent amounts to approximately 35 \% and 8 \% in the presence of 0.1 mM and 0.5 mM \ch{Si}, respectively. After 30 days, all progressions approach full conversion. Assuming the time-dependent amount of 2-line ferrihydrite ($\ch{Fe}(t)$) is proportional to its rate of transformation, 
\begin{equation}
	\frac{d\ch{Fe}(t)}{dt} = -k\ch{Fe}(t),
\end{equation}
can be computed by fitting its mol fraction with respect to the initial iron concentration $\ch{Fe}(t)/\ch{Fe}_0$ to the integrated first order rate equation
$\ch{Fe}(t) = \ch{Fe}_0\times\text{exp}(-kt)$. Fig. \ref{fig:rate_constants} displays the resultant estimated rate constant of phase transformation $k$ in \si{\per\second} as a function of the aqueous \ch{Si} concentration. It is shown that the presence of $1\times10^{-4}$ M of \ch{Si}, corresponding to an initial iron to silicon ratio of 1:200, is sufficient to reduce the rate of corrosion product transformation by one order of magnitude. 
Further reduction is exponential with respect to the silicon concentration and follows
\begin{equation}
	\text{log}_{10}k = \text{log}_{10}k_0 - 14.65 \times[\ch{Si}]^{0.28},
\end{equation}
where $k_0 = (4.04 \pm 1.53)\times10^{-5}$ \si{\per\second} is the rate of 2-line ferrihydrite transformation at pH = 14 in the absence of silicon.
\newpage
\clearpage
\subsection{A suggested mechanism of the influence of Si on the transformation of corrosion products}
\begin{figure}[!ht]
	\centering
	\includegraphics[scale=0.85]{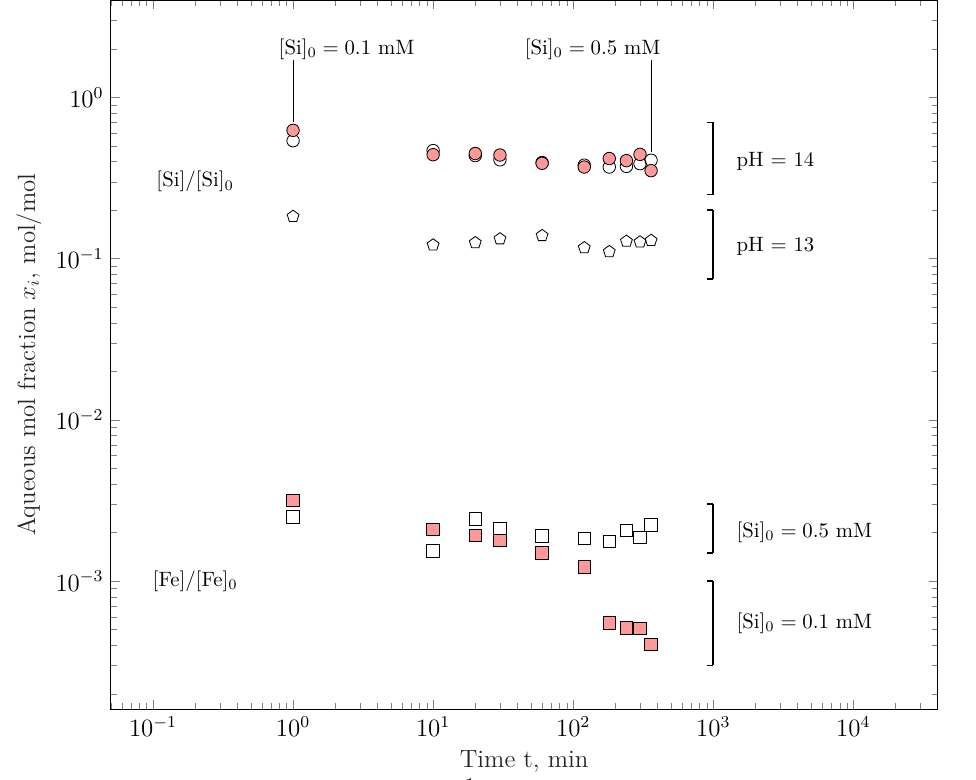}
	\caption{The aqueous mol fraction of iron and silicon with respect to their initial concentrations $[\ch{Fe}]_0$ and $[\ch{Si}]_0$ as a function of time, the pH and the initial silicon concentration. The fraction of aqueous \ch{Si} is approximately constant over time and increases as a function of the aqueous iron concentration and thus, the pH. The progressions of $[\ch{Fe}]/[\ch{Fe}]_0$ decrease as a function of time, where values measured in the presence of $[\ch{Si}]_0 = 0.5$ mM (empty markers) exceed those corresponding to an initial silicon concentration of 0.1 mM (filled markers). Note that the aqueous silicon concentration at pH = 13 and $[\ch{Si}]_0 = $ 0.1 mM is below the LOQ.}
	\label{fig:Fe_Si_mol_fraction}
\end{figure}

XRD (Fig. \ref{fig:XRD_pH14}) and EXAFS (Fig. \ref{fig:EXAFS_R_space}) measurements demonstrate a delayed formation of the crystalline corrosion product goethite from amorphous 2-line ferrihydrite in the presence of silicon. The RDF and X-ray diffractograms of iron hydroxide phases aged in the presence of 0.5 mM \ch{Na2SiO3} are equivalent to the measurements of solid phases stabilised in the presence of 0.1 mM \ch{Na2SiO3} at earlier equilibration times. Changes in the aqueous phase composition over the course of the transformation further indicate that the concentrations of iron are increased in the presence of silicon (compare Figure \ref{fig:ICP}). Vice versa, the fraction of silicon that remains in the aqueous phase with respect to the amount initially dissolved increases w.r.t. the aqueous iron concentration. In either case, the balance, i.e. the difference between the initial and aqueously dissolved number of moles of \ch{Fe} and \ch{Si}, partitions into one of the two solid phases, 2-line ferrihydrite and goethite. These trends are summarised in Fig. \ref{fig:Fe_Si_mol_fraction}.
From the aqueous \ch{Si} concentration measurements presented in Figure \ref{fig:ICP}, it is apparent that the initial reduction in the dissolved silicon concentration is concluded within the first few minutes of equilibration time. As evident from the extended X-ray absorption fine structures presented in Figure \ref{fig:EXAFS_R_space}, the solids stabilised across the same timespan consist exclusively of 2-line ferrihydrite. Consequentially, the difference between the initial and final aqueous silicon concentration must have predominantly associated to 2-line ferrihydrite, rather than goethite. These observations are in agreement with \citeauthor{RN294} \cite{RN294}, demonstrating that the vast majority of dissolved \ch{Si} is associated with 2-line ferrihydrite during its Fe(II)-catalysed transformation to other iron (hydr)oxides at neutral pH. The remaining amount of \ch{Si} that is not adsorbed or taken-up by 2-line ferrihydrite within the first hour of equilibration time appears to influence the morphology of goethite crystals stabilised and the aqueous iron concentration in equilibrium with both solid phases. \citeauthor{RN280} \cite{RN280} show that high levels of \ch{Si} (9.6 mol \% relative to the total number of moles including water) significantly enhances the development of the terminal (021), (111) and (121) crystal faces. Although the aqueous \ch{Si} concentration of 0.1 and 0.5 mM used in this paper is significantly lower, the onset of the (110) and (111) face development over the (002) face is evidenced from an increase in their crystallinity at  0.5 mM (compare Supplementary information, Figure \ref{fig:XRD_peaks}).

\begin{figure}[!ht]
	\centering
	\includegraphics[scale=0.75]{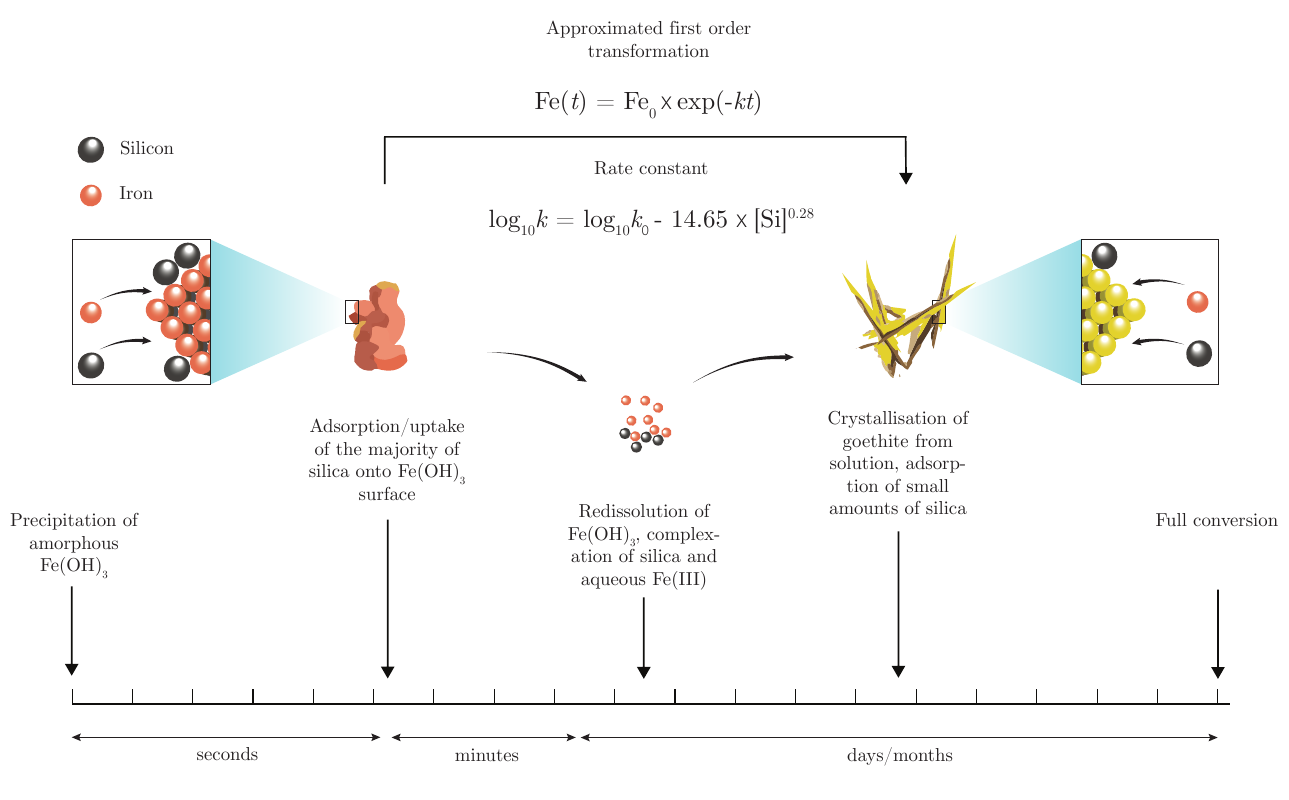}
	\caption{Suggested mechanism of the influence of silicon on the transformation of 2-line ferrihydrite to goethite at alkaline pH. Initially, large quantities of silicon are adsorbed onto 2-line ferrihydrite, the remainder forms aqueous $\ch{Fe}-\ch{Si}$ complexes or adsorbs onto goethite during a later stage in the transformation process. The estimated rate constants $k$, as computed by fitting the molar fractions of 2-line ferrihydrite and goethite to the integrated first order rate equation $\text{Fe}(t) = \text{Fe}_0\times \text{exp}(-kt)$, decrease as a function of the initial aqueous silicon concentration according to 
		the semi-empirical relation $\text{log}_{10}k=\text{log}_{10}k_0 - 14.65\times[\ch{Si}]^{0.28}$, where $k_0 = (4.04 \pm 1.53)\times10^{-5}$ \si{\per\second} is the standard rate constant of transformation at pH = 14 in the absence of silicon.}
	\label{fig:schematic}
\end{figure}

Provided that the adsorption of \ch{Si} onto 2-line ferrihydrite slows down its dissolution\cite{schwertmann2008iron,RN294}, the aqueous iron concentration resulting of the dissolution of 2-line ferrihydrite is expected to be lower in the presence of \ch{Si}. Given additionally that the dissolution of 2-line ferrihydrite limits the rate of goethite precipitation \cite{furcas_speciation_controls_preprint}, while the precipitation of goethite is not occurring instantly as is evidenced by the fact that aqueous iron concentrations plateau above the goethite solubility limit (irrespective of the presence of Si, see Figure \ref{fig:ICP}), the overall aqueous iron concentration as measured by ICP-OES resultant of the combined dissolution of 2-line ferrihydrite and precipitation of goethite is also expected to be lower in the presence of \ch{Si}. As the measured $[\ch{Fe}]$ are instead higher, it is likely that some of the aqueous iron that would otherwise be coordinated as \ch{Fe(OH)4^-} forms an aqueous \ch{Fe}-\ch{Si} complex\cite{Pokrovski_iron_silica_EXAFS}. In comparison to the transformation of 2-line ferrihydrite in the absence of silicon, the observed retardation would thus be due to the competitive formation of goethite and the hitherto unconsidered \ch{Fe}-\ch{Si} complex, effectively reducing the amount of \ch{Fe(OH)4^-} at disposal to precipitate.
The formation of such a complex could explain the strongly increased aqueous iron concentration in cementitious systems containing dissolved \ch{Si} compared to systems aged in the absence of silicon\cite{dilnesa2011iron,dilnesa2014synthesis,furcas_solubility}. Figure \ref{fig:schematic} illustrates these findings in the form of a schematic reaction diagram. 
Further deconvolution of the influence of silicon on the transformation mechanism would require a more detailed investigation of (i) the nature of $\ch{Fe}-\ch{Si}$ complexes formed as well as (ii) the rate of silicon adsorption onto both solid phases and (iii) the rate of 2-line ferrihydrite dissolution as a function of the amount of silicon adsorbed. Moreover, it cannot be excluded that small quantities of \ch{Si} are incorporated into or alter the structure of corrosion products formed. To this effect, the utilised approach to quantify the transformation of 2-line ferrihydrite to goethite by means of XAS LCF may not be applicable at higher silicon concentrations. Instead, the structure and stoichiometry of $\ch{Si}$-bearing iron phases requires rigorous modelling of the local coordination environment of both ligands. 
\subsection{Implications for the durability of reinforced concrete structures}
With regard to the durability of reinforced concrete structures, these findings imply that the pore solution chemistry, i.e. the high degree of alkalinity and the presence of \ch{Si}, maintains an aqueous Fe(III) concentration well above its thermodynamic solubility limit. Irrespective of whether these Fe(III) species are coordinated as one of the Fe(III) hydrolysis products or in the form of an aqueous \ch{Fe}-\ch{Si} complex, they remain mobile in the aqueous phase. Cementitious systems dissolving high amounts of silicon could thus facilitate iron transport and therefore the precipitation of corrosion products distant from the steel-concrete interface. Moreover, the kinetically hindered transformation of 2-line ferrihydrite to goethite could lead to a faster accumulation of internal stresses, as the high-volume corrosion product 2-line ferrihydrite dissolves comparatively slower. As these competing reactive mass-transport phenomena are highly dependent on the local solute concentrations as well as the pore shape in which the corrosion products precipitate, further experimental and computational studies are needed to assess whether the delayed transformation of corrosion products contributes to the structural degradation on a macroscopic level. 

\section{Conclusions}
\label{sec:conclusion}
The findings presented in this study and other research on the precipitation of corrosion products at alkaline pH substantiate the following major conclusions:
\begin{itemize}
	\item The stabilisation of the thermodynamically favourable corrosion product goethite from amorphous 2-line ferrihydrite is delayed in the presence of silicon.
	\item At the aqueous silicon concentration characteristic to a broad range of cementitious systems, the estimated rate constant of transformation decreases exponentially with respect to $[\ch{Si}]$.
	\item The observed retardation of goethite formation is likely a consequence of 3 distinct reactive phenomena, (i) the hindered dissolution of 2-line ferrihydrite, (ii) the formation of a soluble iron-silicon complex and (iii) changes to the morphology of the goethite crystals.
	\item Whilst the influence of \ch{Si} on the dissolution of 2-line ferrihydrite is known \cite{schwertmann2008iron,RN294}, the structure and formation mechanism of the hypothesized \ch{Fe}-\ch{Si} complex as well as the precipitation of goethite in the presence of \ch{Si} must be further researched. 
	\item With regard to the durability of reinforced concrete structures, the observed retardation prospectively generates more internal stresses due to the precipitation of corrosion products, as the specific molar volume of 2-line ferrihydrite is significantly higher than that of goethite. Moreover, the aqueous iron concentration in equilibrium with both phases is higher compared to systems containing no silicon, facilitating the transport of iron complexes across the concrete pore network.
\end{itemize} 

\section*{Acknowledgements}
 All XAS measurements were performed at the PHOENIX beamline at the Swiss Light Source, Paul Scherrer Institut, Villigen, Switzerland. The authors thank Dr. Michael Pl\"{o}tze from the ETH Z\"{u}rich ClayLab at the Institute for Geotechnical Engineering for access to the XRD equipment, help with the measurement and discussion. The authors also thank Raphael Kuhn and Dr. Alexandru Pirvan from the Empa Concrete \& Asphalt Laboratory for their help with the ICP-OES measurements. The authors are further grateful for the financial support provided for Fabio Enrico Furcas by the European Research Council (ERC) under the European Union’s Horizon 2020 research and innovation program (grant agreement no. 848794).
 
\section*{Author contributions}
Fabio E. Furcas: Conceptualization, Investigation, Formal analysis, Visualization, Writing - Original Draft\\
Shishir Mundra: Conceptualization, Investigation, Writing - Review \& Editing, Supervision\\
Barbara Lothenbach: Conceptualization, Formal analysis, Writing - Review \& Editing, Supervision\\
Camelia N. Borca: Investigation, Writing - Review \& Editing\\
Thomas Huthwelker: Investigation, Writing - Review \& Editing\\
Ueli M. Angst: Conceptualization, Writing - Review \& Editing, Supervision, Funding acquisition
\section*{Data availability}
The experimental data generated in this study is openly available in the ETH Research Collection, https://doi.org/10.3929/ethz-b-000658759.
\newpage
\clearpage
\section*{Supplementary information} \label{sec:appendix}
\setcounter{table}{0}
\renewcommand{\thetable}{S\arabic{table}}
\setcounter{figure}{0}
\renewcommand{\thefigure}{S\arabic{figure}}
\begin{figure}[!ht]
	\centering
	%\hspace{-25pt}
	\begin{subfigure}[b]{0.5\textwidth}
		\centering
		\includegraphics[scale=0.77]{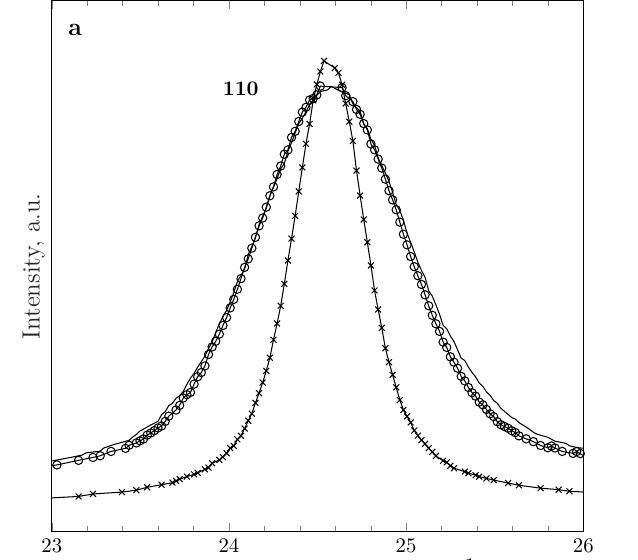}
	\end{subfigure}%
	%\vskip\baselineskip
	\begin{subfigure}[b]{0.5\textwidth}
		\hspace{0.2cm}
		\centering
		\includegraphics[scale=0.77]{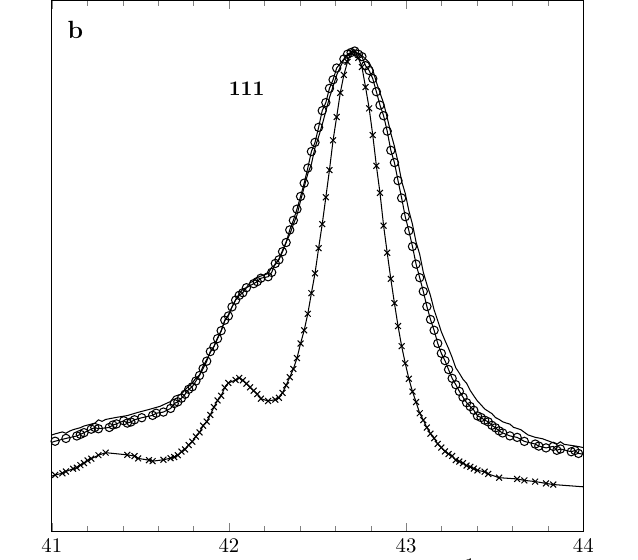}
	\end{subfigure}
	\vskip\baselineskip
	\begin{subfigure}[b]{0.5\textwidth}
		\centering
		\includegraphics[scale=0.77]{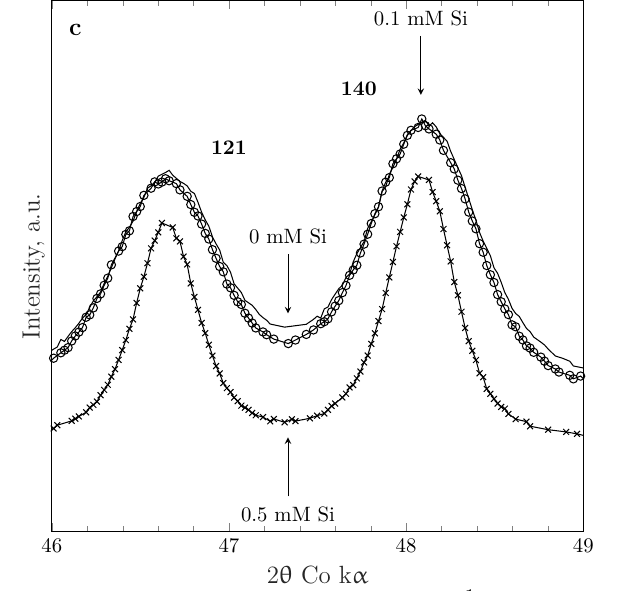}
	\end{subfigure}%
	%\vskip\baselineskip
	\begin{subfigure}[b]{0.5\textwidth}
		\hspace{0.2cm}
		\centering
		\includegraphics[scale=0.77]{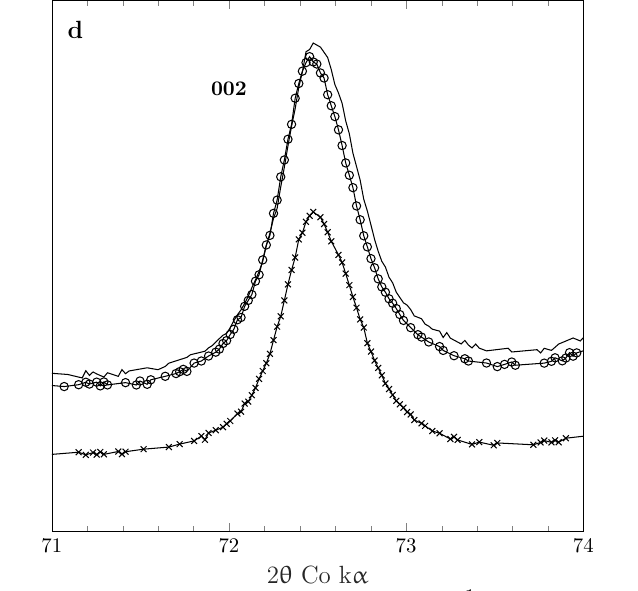}
	\end{subfigure}
	\caption{X-ray diffractogram 110 (Fig. \ref{fig:XRD_peaks}a), 111 (Fig. \ref{fig:XRD_peaks}b), 121, 140 (Fig. \ref{fig:XRD_peaks}c) and 002 (Fig. \ref{fig:XRD_peaks}d) peaks of the iron hydroxide powders extracted from supersaturated stock solution after 30 days the presence of no silica (continuous line), 0.1 mM \ch{Na2SiO3} (circular marker) and 0.5 mM \ch{Na2SiO3} (cross marker).}
	\label{fig:XRD_peaks}
\end{figure}
\newpage
\clearpage
\begin{figure}[!ht]
	\centering
	\begin{subfigure}[b]{\textwidth}
		\includegraphics[scale=0.85]{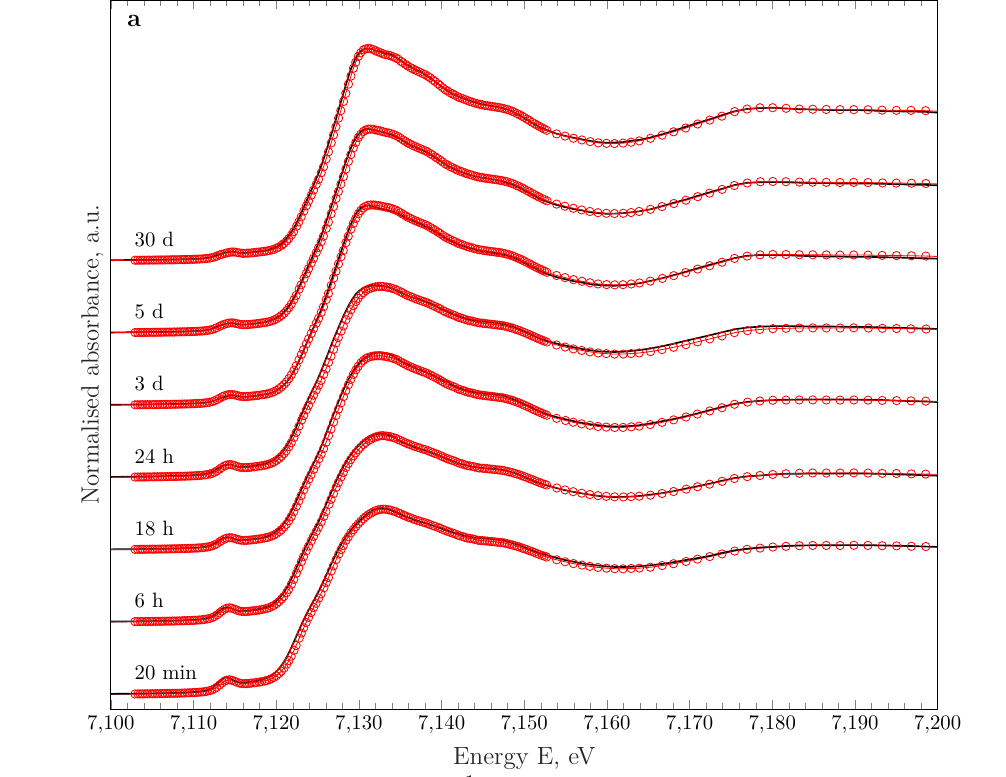}
	\end{subfigure}%
	\vskip\baselineskip
	\begin{subfigure}[b]{\textwidth}
		\includegraphics[scale=0.85]{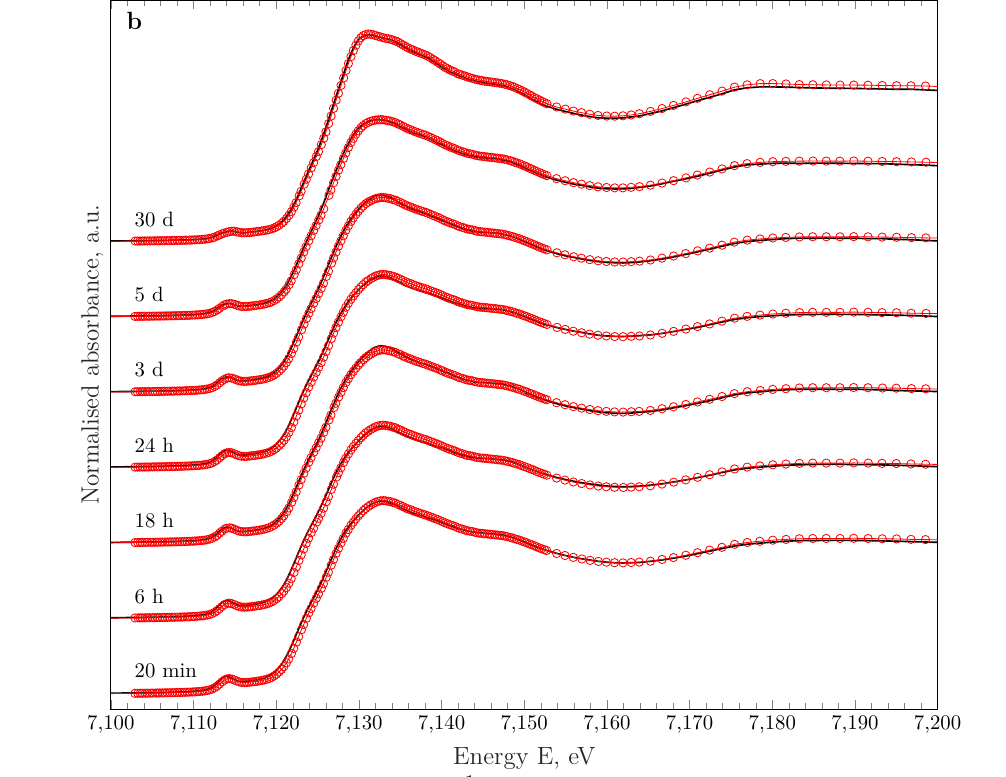}
	\end{subfigure}
	\caption{Normalised Fe K-edge XANES spectra of iron hydroxide powders (continuous lines) extracted from supersaturated stock solutions at pH = 14 containing 20.0 mM $\ch{FeCl3}\cdot6\ch{H2O(cr)}$ and 0.1 mM \ch{Na2SiO3} (Fig. \ref{fig:XANES_LCF}a) and 0.5 mM \ch{Na2SiO3} (Fig. \ref{fig:XANES_LCF}b), together with their linear combination fits (circular markers). Spectra were fitted using the reference standards 2-line ferrihydrite and goethite in the XANES region from 20 eV below to 40 eV above the absorption edge.}
	\label{fig:XANES_LCF}
\end{figure}
\newpage
\clearpage
\begin{figure}[!ht]
	\centering
	\begin{subfigure}[b]{\textwidth}
		\includegraphics[scale=0.85]{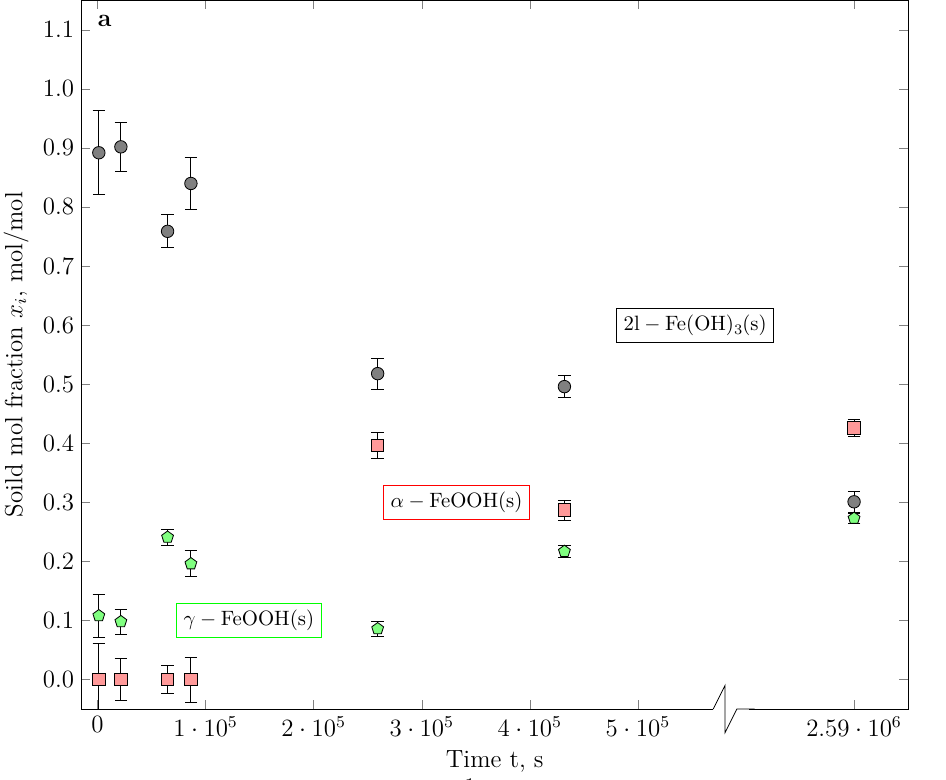}
	\end{subfigure}%
	\vskip\baselineskip
	\begin{subfigure}[b]{\textwidth}
		\includegraphics[scale=0.85]{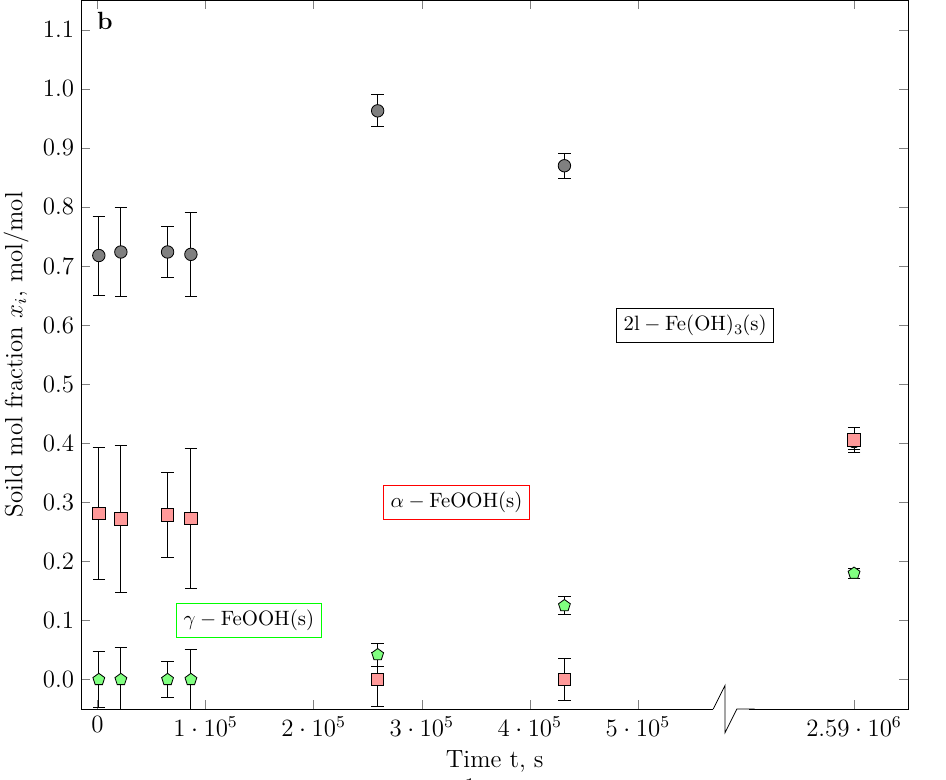}
	\end{subfigure}
	\caption{Time dependent solid molar fractions of the reference standards 2-line ferrihydrite, lepidocrocite and goethite in the presence of 0.1 mM \ch{Na2SiO3} (Fig. \ref{fig:XANES_LCF_progression_lp}a) and 0.5 mM \ch{Na2SiO3} (Fig. \ref{fig:XANES_LCF_progression_lp}b) at pH = 14. The data is obtained from LCF of the XANES spectra displayed in Fig. \ref{fig:XANES_LCF}.}
	\label{fig:XANES_LCF_progression_lp}
\end{figure}
\newpage
\clearpage
\begin{table}[ht]
	\scriptsize
	\setlength\extrarowheight{7pt}
	\centering
	\caption{LOD and LOQ in \si{\micro\gram\per\liter} and \si{\micro\mole\per\liter} for the low-concentration element Fe. Concentrations are obtained by means of a linear 8-point interpolation containing 0.01, 0.10, 0.50, 1.00, 5.00, 10.00, 25.00 and 50.00 ppm of the measured elements.}
	\label{tab:ICP_LOD_LOQ}
	\resizebox{\textwidth}{!}{
		\begin{tabular}{l l l p{0.1cm} r r p{0.1cm} r r}
			\hline\hline
			Element & Spectral line, nm & SD & {} & \multicolumn{2}{l}{LOD} & {} & \multicolumn{2}{l}{LOQ} \\ \cmidrule{3-3} \cmidrule{5-6} \cmidrule{8-9}
			{} & {} & (ppm) & {} & (\si{\micro\gram\per\liter}) & (\si{\micro\mole\per\liter}) & {} & (\si{\micro\gram\per\liter}) & (\si{\micro\mole\per\liter}) \\
			\hline
			\ch{Fe} & 234.350 & $2.961\cdot10^{-4}$ & {} & 0.888 & 0.016 & {} & 2.961 & 0.053 \\
			{} & 238.204 & $4.726\cdot10^{-4}$ & {} & 1.418 & 0.025 & {} & 4.726 & 0.085 \\
			{} & 239.563 & $4.348\cdot10^{-4}$ & {} & 1.304 & 0.023 & {} & 4.348 & 0.078 \\
			{} & 259.940 & $9.081\cdot10^{-4}$ & {} & 2.724 & 0.049 & {} & 9.081 & 0.163 \\
			\hline
			\ch{Si} & 212.412 & $2.739\cdot10^{-3}$ & {} & 8.215 & 0.293 & {} & 27.385 & 0.975 \\
			{} & 250.690 & $3.102\cdot10^{-3}$ & {} & 9.307 & 0.331 & {} & 31.023 & 1.150 \\
			{} & 251.611 & $2.224\cdot10^{-3}$ & {} & 6.672 & 0.238 & {} & 22.204 & 0.792 \\
			{} & 288.158 & $2.639\cdot10^{-3}$ & {} & 7.918 & 0.282 & {} & 26.394 & 0.940 \\
			\hline\hline
		\end{tabular}
	}
\end{table}%
\begin{figure}[!ht]
	\centering
	\hspace{-50pt}
	\begin{subfigure}[b]{0.48\textwidth}
		\centering
		\includegraphics{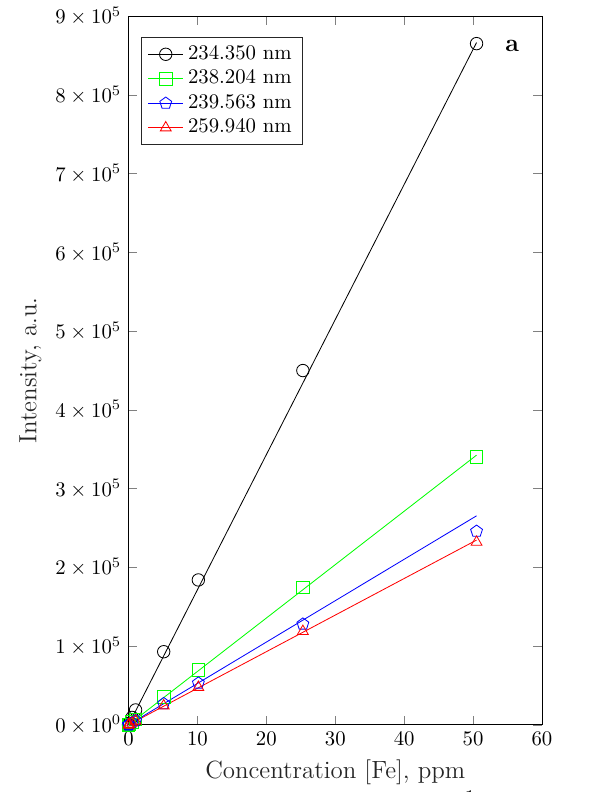}
	\end{subfigure}%
	\hspace{40pt}
	%\vskip\baselineskip
	\begin{subfigure}[b]{0.48\textwidth}
		%\hspace{0.2cm}
		\centering
		\includegraphics{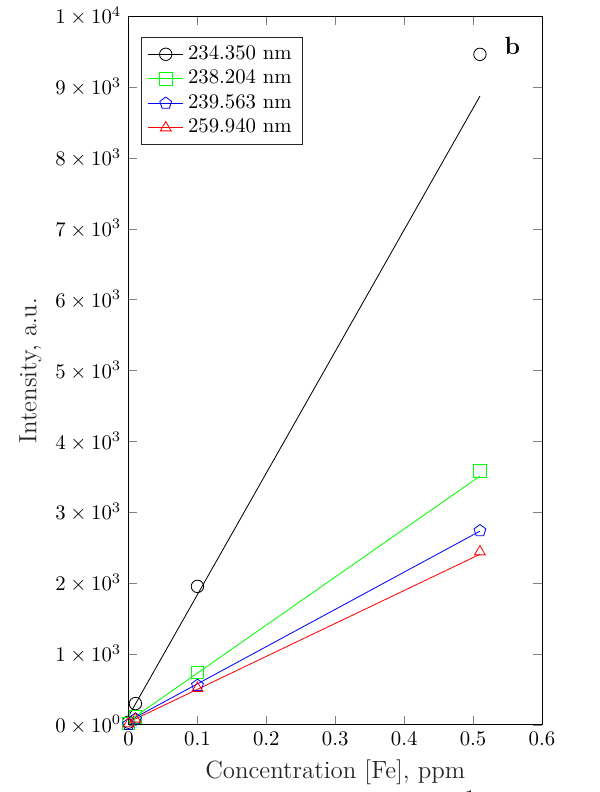}
	\end{subfigure}
	\caption{Inductively coupled plasma optical emission spectroscopy calibration for the element iron, corresponding to the solution composition displayed in Table \ref{tab:ICP_Standards_noSi}. Figure \ref{fig:ICP_calibration_Fe_8}a displays the full calibration range (0.01, 0.05, 0.10, 0.50, 1.00, 2.50, and 5.00 ppm) and Figure \ref{fig:ICP_calibration_Fe_8}b shows a zoomed-in section of the three lowest standards above the blank (0.01, 0.05 and 0.10 ppm). All linear fits feature a coefficient of determination R\textsuperscript{2}$\geq$0.999.}
	\label{fig:ICP_calibration_Fe_8}
\end{figure}
\begin{figure}[!ht]
	\centering
	\hspace{-50pt}
	\begin{subfigure}[b]{0.48\textwidth}
		\centering
		\includegraphics{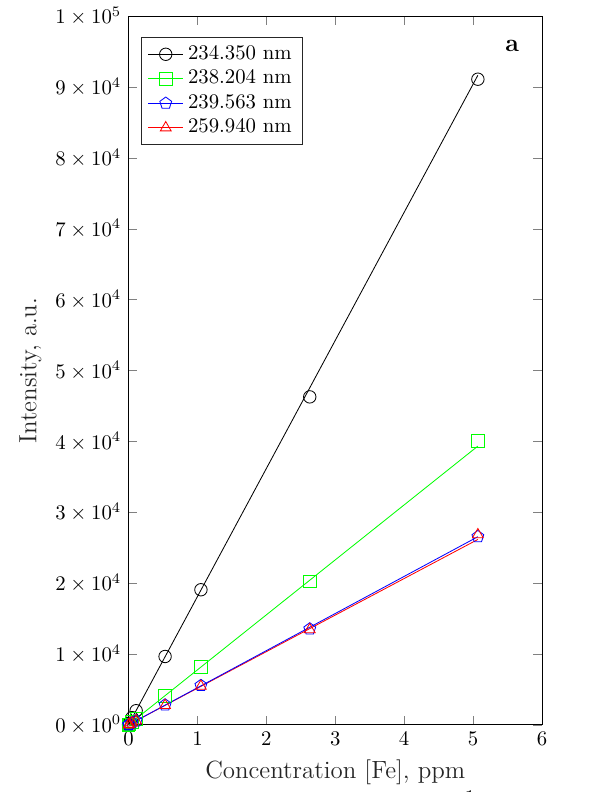}
	\end{subfigure}%
	\hspace{40pt}
	%\vskip\baselineskip
	\begin{subfigure}[b]{0.48\textwidth}
		%\hspace{0.2cm}
		\centering
		\includegraphics{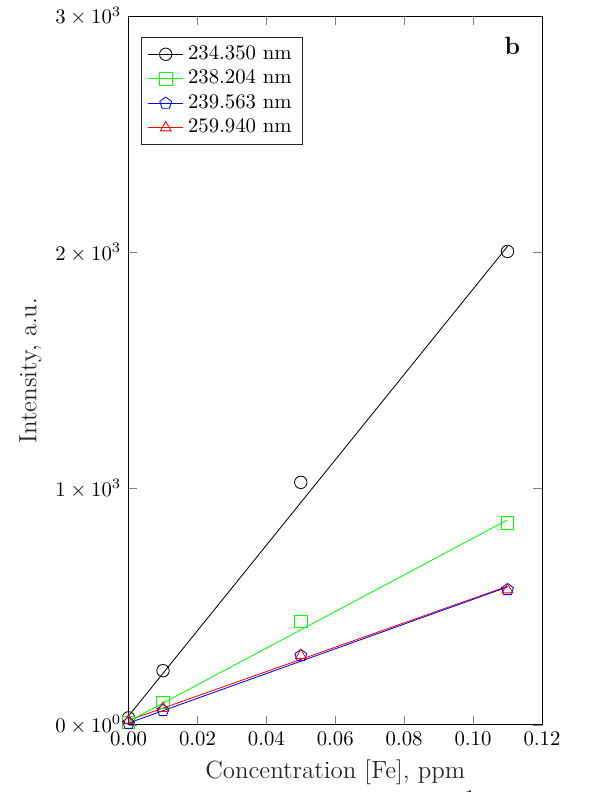}
	\end{subfigure}
	\caption{Inductively coupled plasma optical emission spectroscopy calibration for the element iron, corresponding to the solution composition displayed in Table \ref{tab:ICP_Standards_Si}. Figure \ref{fig:ICP_calibration_Fe}a displays the full calibration range (0.01, 0.05, 0.10, 0.50, 1.00, 2.50, and 5.00 ppm) and Figure \ref{fig:ICP_calibration_Fe}b shows a zoomed-in section of the three lowest standards above the blank (0.01, 0.05 and 0.10 ppm). All linear fits feature a coefficient of determination R\textsuperscript{2}$\geq$0.999.}
	\label{fig:ICP_calibration_Fe}
\end{figure}
\begin{figure}[!ht]
	\centering
	\hspace{-50pt}
	\begin{subfigure}[b]{0.48\textwidth}
		\centering
		\includegraphics{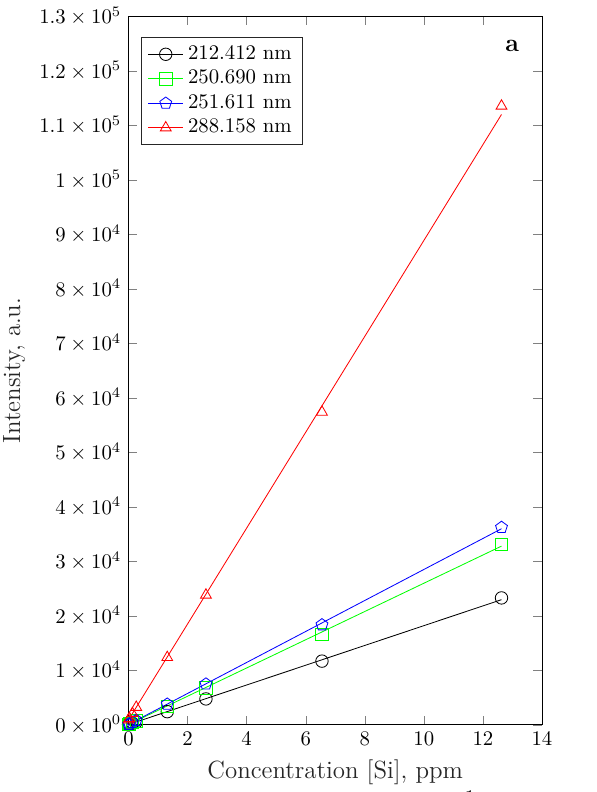}
	\end{subfigure}%
	\hspace{40pt}
	%\vskip\baselineskip
	\begin{subfigure}[b]{0.48\textwidth}
		%\hspace{0.2cm}
		\centering
		\includegraphics{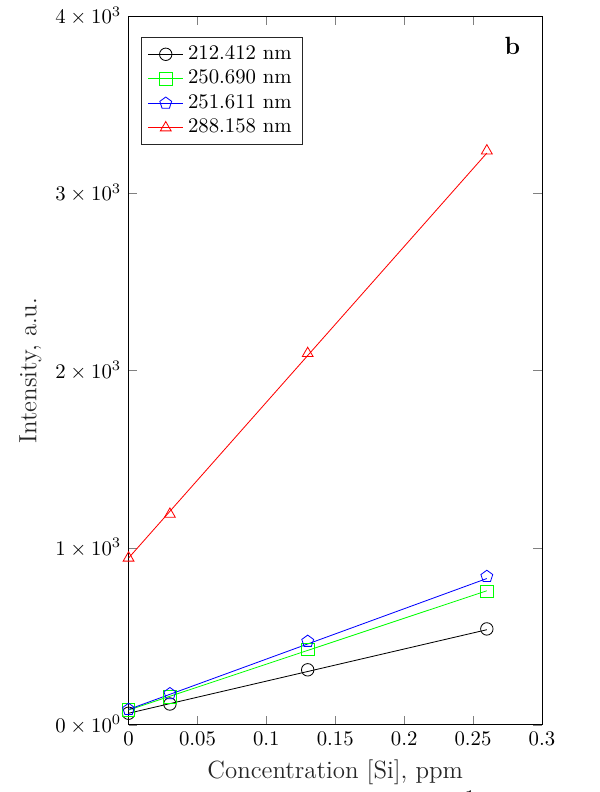}
	\end{subfigure}
	\caption{Inductively coupled plasma optical emission spectroscopy calibration for the element silicon, corresponding to the solution composition displayed in Table \ref{tab:ICP_Standards_Si}. Figure \ref{fig:ICP_calibration_Si}a displays the full calibration range (0.02, 0.12, 0.25, 1.25, 2.50, 6.50, and 12.50 ppm) and Figure \ref{fig:ICP_calibration_Si}b shows a zoomed-in section of the three lowest standards above the blank (0.02, 0.12 and 0.25 ppm). All linear fits feature a coefficient of determination R\textsuperscript{2}$\geq$0.999.}
	\label{fig:ICP_calibration_Si}
\end{figure}
\newpage
\clearpage
\begin{table}[ht]
	\scriptsize
	\setlength\extrarowheight{5pt}
	\centering
	\caption{Elemental composition of the ICP calibration solutions used to determine the total amount dissolved aqueous iron at pH = 13 and 14 in the absence of silica. For all runs, the blank solution consisted of 2 wt.\% \ch{HNO3} in UPW.}
	\label{tab:ICP_Standards_noSi}
	\resizebox{\textwidth}{!}{
		\begin{tabular}{l p{0.1cm} r p{0.1cm} r p{0.1cm} r p{0.1cm} r p{0.1cm} r p{0.1cm} r p{0.1cm} r p{0.1cm} r p{0.1cm} r}
			\hline\hline
			{Standard} & {} & \multicolumn{17}{c}{Element, ppm}\\
			{Number} {} & & \multicolumn{1}{c}{Na} & {} & \multicolumn{1}{c}{K} & {} & \multicolumn{1}{c}{Ca} & {} & \multicolumn{1}{c}{Mg} & {} & \multicolumn{1}{c}{Al} & {} & \multicolumn{1}{c}{Fe} & {} & \multicolumn{1}{c}{S} & {} & \multicolumn{1}{c}{Si} & {} & \multicolumn{1}{c}{P} \\
			\cmidrule{1-1} \cmidrule{3-3} \cmidrule{5-5} \cmidrule{7-7} \cmidrule{9-9} \cmidrule{11-11} \cmidrule{13-13} \cmidrule{15-15} \cmidrule{17-17} \cmidrule{19-19}\\
			8 & {} & 50.5506 & {} & 50.2203 & {} & 50.3531 & {} & 50.4780 & {} &	20.2052 & {} & 50.3020 & {} & 50.2344 & {} & 50.3595 & {} & 50.3269 \\		
			7 & {} & 25.3110	& {} & 25.1456 & {} & 25.2121 & {}	& 25.2747 & {} & 10.1169 & {}	& 25.1865 & {} &	25.1527 & {} & 25.2153 & {} &	25.1990 \\
			6 & {} & 10.1346 & {}	& 10.0684 & {} & 10.0950 & {}	& 10.1201 & {} &	4.0508 & {}	& 10.0848 & {} &	10.0712 & {}	& 10.0963 & {} &	10.0898 \\
			5 & {} & 5.0766 & {} & 5.0435 & {} & 5.0568 & {} & 5.0693 & {} & 2.0291 & {} & 5.0517 & {} & 5.0449 & {} & 5.0574 & {} & 5.0542 \\
			4 & {} & 1.0239 & {} & 1.0172 & {} & 1.0199 & {} & 1.0224 & {} & 0.4093 & {} & 1.0188 & {} & 1.0175 & {} & 1.0200 & {} & 1.0194 \\
			3 & {} & 0.5096 & {} & 0.5063 & {} & 0.5076 & {} & 0.5089 & {} & 0.2037 & {} & 0.5071 & {} & 0.5064 & {} & 0.5077 & {} & 0.5640 \\
			2 & {} & 0.1017 & {} & 0.1010 & {} & 0.1013 & {} & 0.1016 & {} & 0.0407 & {} & 0.1012 & {} & 0.1011 & {} & 0.1013 & {} & 0.1013 \\
			1 & {} & 0.0101 & {} & 0.0101 & {} & 0.0101 & {} & 0.0102 & {} & 0.0041 & {} & 0.0101 & {} & 0.0101 & {} & 0.0101 & {} & 0.0101 \\
			\hline\hline
		\end{tabular}
	}
\end{table}%
\begin{table}[ht]
	\scriptsize
	\setlength\extrarowheight{5pt}
	\centering
	\caption{Elemental composition of the ICP calibration solutions used to determine the total amount dissolved aqueous iron and silica at pH = 13 and 14 for $[\ch{Si}]=1\cdot10^{-4}$ M and $[\ch{Si}]=5\cdot10^{-4}$ M. For all runs, the blank solution consisted of 2 wt.\% \ch{HNO3} in UPW.}
	\label{tab:ICP_Standards_Si}
	\resizebox{\textwidth}{!}{
		\begin{tabular}{l p{0.1cm} r p{0.1cm} r p{0.1cm} r p{0.1cm} r p{0.1cm} r p{0.1cm} r p{0.1cm} r p{0.1cm} r p{0.1cm} r}
			\hline\hline
			{Standard} & {} & \multicolumn{17}{c}{Element, ppm}\\
			{Number} {} & & \multicolumn{1}{c}{Na} & {} & \multicolumn{1}{c}{K} & {} & \multicolumn{1}{c}{Ca} & {} & \multicolumn{1}{c}{Mg} & {} & \multicolumn{1}{c}{Al} & {} & \multicolumn{1}{c}{Fe} & {} & \multicolumn{1}{c}{S} & {} & \multicolumn{1}{c}{Si} & {} & \multicolumn{1}{c}{P} \\
			\cmidrule{1-1} \cmidrule{3-3} \cmidrule{5-5} \cmidrule{7-7} \cmidrule{9-9} \cmidrule{11-11} \cmidrule{13-13} \cmidrule{15-15} \cmidrule{17-17} \cmidrule{19-19}\\
			7 & {} & 50.4817 & {} & 50.1734 & {} & 49.6668 & {} & 20.2355 & {} & 20.2262 & {} & 5.0717 & {} & 50.3728 & {} & 12.6314 & {} & - \\		
			6 & {} & 26.1642 & {} & 26.0044 & {} & 25.7418 & {}	& 10.4878 & {} & 10.4830 & {}& 2.6286 & {} 	&	26.1077 & {} & 6.5467 & {} &	- \\
			5 & {} & 10.4703 & {}	& 10.4064 & {} & 10.3013 & {} & 4.1970 & {} & 4.1951 & {}& 1.0519 & {} 	&	10.4477 & {}& 2.6198 & {} &	- \\
			4 & {} & 5.2400 & {} & 5.2080 & {} & 5.1555 & {} & 2.1005 & {} & 2.0995 & {} & 0.5264 & {} 		& 5.2287 & {} 	& 1.3111 & {} & - \\
			3 & {} & 1.0496 & {} & 1.0432 & {} & 1.0327 & {} & 0.4207 & {} & 0.4206 & {} & 0.1055 & {} 		& 1.0474 & {} 	& 0.2626 & {} & - \\
			2 & {} & 0.5293 & {} & 0.5261 & {} & 0.5208 & {} & 0.2122 & {} & 0.2121 & {} & 0.0532 & {} 		& 0.5282 & {} 	& 0.1324 & {} & - \\
			1 & {} & 0.1057 & {} & 0.1051 & {} & 0.1040 & {} & 0.0424 & {} & 0.0424 & {} & 0.0106 & {} 		& 0.1055 & {} 	& 0.0265 & {} & - \\
			\hline\hline
		\end{tabular}
	}
\end{table}%
\newpage
\clearpage
\bibliographystyle{unsrtnat}
\bibliography{Bibliography}

\end{document}